\documentclass[12pt]{spieman}  
\usepackage{amsmath,amsfonts,amssymb}
\usepackage{graphicx}
\usepackage{setspace}
\usepackage{tocloft}

\usepackage{lineno}

\title{Demonstration of the broadband half-wave plate using the nine-layer sapphire for the CMB polarization experiment}

\author[a]{Kunimoto Komatsu}
\author[b]{Tomotake Matsumura}
\author[c]{Hiroaki Imada}
\author[a]{Hirokazu Ishino}
\author[b]{Nobuhiko Katayama}
\author[b]{Yuki Sakurai}
\affil[a]{Okayama University, Okayama 700-8530, Japan}
\affil[b]{Kavli Institute for the Physics and Mathematics of the Universe (WPI), University of Tokyo, Kashiwa 277-8583, Japan}
\affil[c]{Laboratoire de l'Acc\'{e}l\'{e}rateur Lin\'{e}aire, Universit\'{e} Paris-Sud, CNRS/IN2P3, Universit\'{e} Paris-Saclay, Orsay, France}

\cftpagenumbersoff{figure}
\cftpagenumbersoff{table} 

\begin{document} 
\maketitle

\begin{abstract}
We report the development of the achromatic half-wave plate (AHWP) at millimeter wave for cosmic microwave background polarization experiments.
We fabricate an AHWP consisting of nine A-cut sapphire plates based on the Pancharatnam recipe to cover a wide frequency range.
The modulation efficiency and the phase are measured in a frequency range of 33 to 260~GHz with incident angles up to 10 degrees.
We find the measurements at room temperature are in good agreement with the predictions. 
This is the most broadband demonstration of an AHWP at millimeter wave.
\end{abstract}

\keywords{CMB polarization, B-mode, Achromatic half-wave plate, Space mission, Polarimetry, Millimeter wave}

{\noindent \footnotesize\textbf{*}Kunimoto Komatsu,  \linkable{k.komatsu@s.okayama-u.ac.jp} }


\section{INTRODUCTION}
\label{sec:intro}  
\par Cosmic inflation is one of the theoretical models that give rise to the initial conditions of the hot big-bang in our universe. 
The rapid space expansion, immediately after the beginning of the universe, produced quantum fluctuations in space-time.
Thus, it generated primordial gravitational waves, which imprinted the B-mode polarization in the cosmic microwave background (CMB). 
The strength of the gravitational waves is represented by a tensor-to-scalar ratio $r$. The majority of the single-scalar-field slow-roll inflation models predict the value of $r$ to be $> 0.01$. 
The target accuracy for future CMB polarization measurements is set to be less than 0.001~\cite{litebird,cmc-s4}.

The required level of the B-mode polarization measurements for the primordial gravitational waves is on the order of a nano-Kelvin corresponding to $r\sim 0.001$; a precise control of systematic effects is needed.
A main source of systematic effects in experiments that rely on pair-differencing, is caused by the different characteristics in a mutually orthogonal pair of detectors in the polarization sensitive orientation.
The $1/f$ noise in the measurement system and from atmospheric fluctuation is also a source of the contamination in the large angular scales, where the primordial gravitational wave signal is prominent.
The measurements with a polarization modulation employing a rotating half-wave plate (HWP) can mitigate those systematics.
Another challenge is to separate the CMB polarization with respect to the polarized emission from our own Galaxy.
We need to subtract more than 99\% of the polarized foreground emission to achieve the target precision of $r$. 
The standard way to differentiate the foreground emission and the CMB is to make use of the difference in frequency spectra of the sources;
the CMB is known to have a perfect blackbody while the foreground emission, such as the synchrotron and dust emissions, have spectra different from the CMB.
In order to measure the difference of the spectrum shape, we need an optical system with broad frequency coverage.

In the past, the HWP was first implemented to the CMB experiment by MAXIPOL~\cite{maxipol}, and has been followed by a number of CMB experiments, including ABS, EBEX, SPIDER, POLARBEAR~\cite{abs ,ebex, spider, polarbear}.
Upcoming experiments also plan to employ a similar system~\cite{ysakurai_spie,charles_spie,so,swipe}. 
The HWP is made of a birefringent material plate with an optic axis\cite{hecht} parallel to the surface.
When the thickness of the plate is chosen properly, the phase difference between the ordinary and extraordinary electric waves passing through the plate becomes $\pi$ radians, i.e., a half wavelength.
When the HWP is rotated with respect to a linear polarization-sensitive detector (in Figure~\ref{fig:signal_meas}),
the incident plane of polarization rotates at a rate of twice the HWP rotation angle, and the measured intensity by the detector appears at a rate of four times the HWP angle.
The continuously rotating HWPs modulate the signal at four times the rotational frequency of the HWP. 
As a result, we can reconstruct linear polarization components $Q$ and $U$ from the signal modulated at four times the rotational frequency of a HWP with a single detector.
Correspondingly, the requirement to match the detector properties between two detectors is greatly relaxed. 

The retardance of a waveplate can be written as $\delta=2\pi \frac{\Delta n d}{\lambda}$, where $\Delta n = |n_{\mathrm{e}}-n_{\mathrm{o}}|$ is the difference between the indices $n_{\mathrm{o}}$ and $n_{\mathrm{e}}$ for the ordinary and extraordinary rays, respectively, $d$ is the thickness, and $\lambda$ is the wavelength. 
The single HWP, which is made of a birefringent material plate, can be used only at the specific wavelength and its harmonics determined by the material and thickness. 
While the single HWP is generally a single-frequency device, Pancharatnam proposed to stack multiple wave plates to broaden the frequency range~\cite{pancharatnam_1,pancharatnam_2}. 

In this paper, we describe our prototype design of the Pancharatnam-based achromatic HWP (AHWP) composed of nine sapphire plates for use in the CMB polarimetry. 
The design and the experimental demonstration appear in Hanany et al. (2005),  Savini et al. (2006), and Pisano et al. (2006)~\cite{hanany,savini,pisano}.
We designed and constructed the prototype AHWP and evaluated it experimentally in a millimeter wave band, from 33 to 260~GHz, which is the widest demonstrated bandwidth at millimeter-wave.
We discussed the results including all the features which we have observed in the measured modulation efficiency.
This development is motivated to develop a broadband HWP for the next-generation CMB polarization satellite, LiteBIRD~\cite{litebird,litebird_new}.
The initial design of the observational frequency band of the low-frequency telescope (LFT) was from 34 to 270~GHz, and thus we aim for this range as a development goal.
After the design iteration, now LiteBIRD LFT covers from 34 to 161~GHz~\cite{litebird_new}.
As a result, we present the results of the development that cover the wider range as compared to the current LiteBIRD LFT frequency coverage.
And in this paper, when referring to the frequency range or band of LiteBIRD, it refers to the old one.

\section{FORMALISM}
\label{sec:form}  
\begin{figure}[t]
\begin{center}
\includegraphics[width=\hsize]{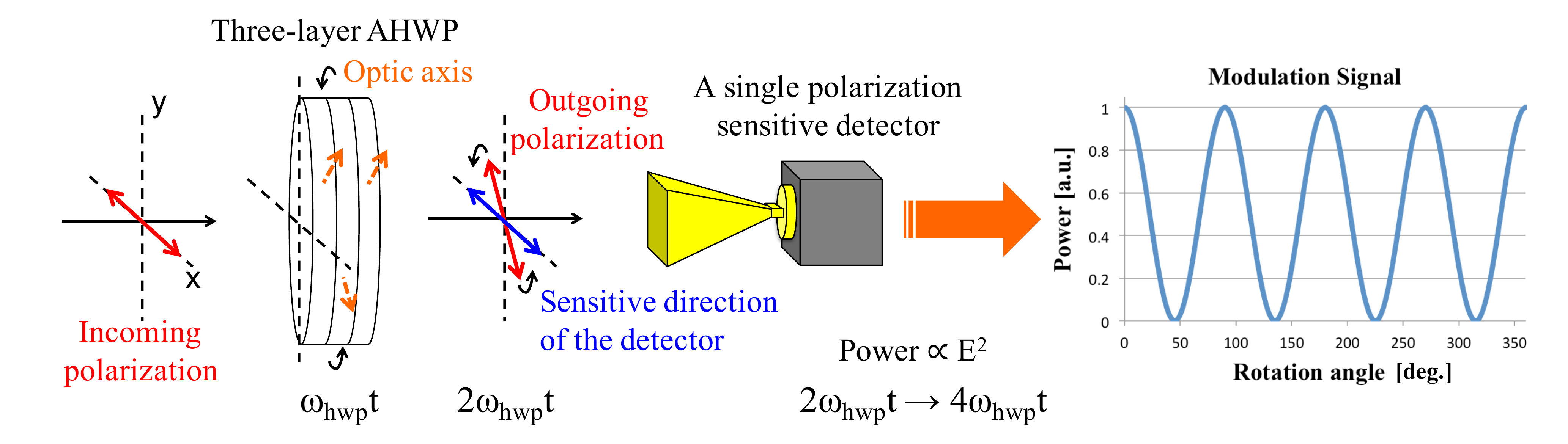}
\caption{\label{fig:signal_meas} A conceptual sketch of the three-layer AHWP polarimetry. The linearly polarized plane waves propagate from left to right in this figure. An example of the modulated signal resulting from one revolution of the HWP is shown in the most right hand side. The amplitude is related to the polarized intensity and the phase is related to the polarization angle of the incident radiation.}
\end{center}
\end{figure}

We construct the system in Figure~\ref{fig:signal_meas}, where the power of polarized light after passing through a continuous rotating AHWP is measured by a single polarization sensitive detector.
The polarization state is expressed by using Stokes vectors and Mueller matrices as,
\begin{equation}
S_{\rm out}=GR(-\omega_{\rm hwp} t)\Gamma_{\rm AHWP} R(\omega_{\rm hwp} t)S_{\rm in},
\label{eq:Sout}
\end{equation}

\begin{equation}
R(\rho)=\left(
\begin{array}{cccc}
1&0&0&0\\
0&\cos2\rho&-\sin2\rho&0 \\
0&\sin2\rho&\cos2\rho&0 \\
0&0&0&1 
\end{array}
\right),
\label{mt:rotation_mueller}
\end{equation}

\begin{equation}
G=\frac{1}{2}\left(
\begin{array}{cccc}
1&1&0&0\\
1&1&0&0 \\
0&0&0&0 \\
0&0&0&0 
\end{array}
\right),
\label{mt:polarizer}
\end{equation}

\begin{equation}
\Gamma_{\rm AHWP}=\left(
\begin{array}{cccc}
M_{\rm II}&M_{\rm IQ}&M_{\rm IU}&M_{\rm IV} \\
M_{\rm QI}&M_{\rm QQ}&M_{\rm QU}&M_{\rm QV} \\
M_{\rm UI}&M_{\rm UQ}&M_{\rm UU}&M_{\rm UV} \\
M_{\rm VI}&M_{\rm VQ}&M_{\rm VU}&M_{\rm VV} 
\end{array}
\right),
\label{mt:mueller_mt}
\end{equation}
where $S_{\rm in}=(I_{\rm in}, Q_{\rm in}, U_{\rm in}, V_{\rm in})$ and $S_{\rm out}=(I_{\rm out}, Q_{\rm out}, U_{\rm out}, V_{\rm out})$ are the Stokes vectors of the incident and outgoing radiation. $\Gamma_{\rm AHWP}$ is the Mueller matrix of an AHWP, $R$ is the rotation matrix, $\rho$ is the rotation angle of the AHWP with respect to the detector coordinate, $G$ is Mueller matrix for a polarizer which defines the polarization-sensitive orientation of the detector, $t$ is time, and $\omega_{\rm hwp}$ is the angular frequency of the AHWP rotation.

The circular polarization of the CMB is negligible. 
The circular polarization level for the Galactic synchrotron emission is estimated to be about 10 nK at most for 10 GHz frequency~\cite{king2016} and about 4 pK at 100GHz assuming the frequency dependence of a power of -3.5 in the Rayleigh-Jeans units (a factor of 1.29 is different between Rayleigh-Jeans units and blackbody temperature at 100GHz).  The circular polarization due to the dust is considered to be even lower~\cite{Montero_Camacho_2018}. 
In summary, the circular polarization from foreground components is negligible compared to the CMB B-mode polarization of about a few nK.
Therefore, we set $V_{\rm in}=0$ in this paper.

For the normal incidence, the detected signal, $I_{\rm out}$, can be written as a function of time as:
\begin{equation}
\begin{split}
I_{\rm out}(t)=&D_{\rm 0I}I_{\rm in}+D_{\rm 0Q}Q_{\rm in}+D_{\rm 0U}U_{\rm in} \\ 
&+D_{\rm 2I}I_{\rm in}\cos(2\omega_{\rm hwp} t-2\phi_{\rm 0})+D_{\rm 2}\sqrt{Q_{\rm in}^{2}+U_{\rm in}^{2}}\cos(2\omega_{\rm hwp} t-2\phi_{2}) \\
&\qquad \qquad \qquad \qquad \qquad \qquad +D_{\rm 4}\sqrt{Q_{\rm in}^{2}+U_{\rm in}^{2}}\cos(4\omega_{\rm hwp} t-4\phi_{\rm 4}), \\
\label{eq:Iout}
\end{split}
\end{equation}
where,
\begin{equation*}
\begin{split}
D_{\rm 0I}&=\frac{1}{2}M_{\rm II} \\
D_{\rm 0Q}&=\frac{1}{4}(M_{\rm QQ}+M_{\rm UU}) \\
D_{\rm 0U}&=\frac{1}{4}(M_{\rm QU}-M_{\rm UQ}) \\
D_{\rm 2I}&=\frac{1}{2}\sqrt{M_{\rm UI}^{2}+M_{\rm QI}^{2}} \\
\phi_{\rm 0}&=\frac{1}{2}\arctan\frac{M_{\rm UI}}{M_{\rm QI}}  \\
D_{\rm 2}&=\frac{1}{2}\sqrt{M_{\rm IQ}^{2}+M_{\rm IU}^{2}} \\
\phi_{\rm 2}&=\frac{1}{2}\arctan\frac{M_{\rm IU}}{M_{\rm IQ}} +\frac{1}{2}\arctan\frac{U_{\rm in}}{Q_{\rm in}} \\
D_{\rm 4}&=\frac{1}{4}\sqrt{(M_{\rm QQ}-M_{\rm UU})^{2}+(M_{\rm QU}+M_{\rm UQ})^{2}}\\
\phi_{\rm 4}&=\frac{1}{4}\arctan\frac{M_{\rm QU}+M_{\rm UQ}}{M_{\rm QQ}-M_{\rm UU}}+\frac{1}{4}\arctan\frac{U_{\rm in}}{Q_{\rm in}}. \\
\end{split}
\end{equation*}
From the demodulation at $4\omega_{\rm hwp}$, we can extract the polarization power of the incident light.
In order to demonstrate the performance of the AHWP and to compare it with the model, we define the modulation efficiency, $\epsilon$, as:
\begin{equation}
\epsilon=\frac{D_{\rm 4}\sqrt{Q_{\rm in}^{2}+U_{\rm in}^{2}}}{D_{\rm 0I}I_{\rm in}+D_{\rm 0Q}Q_{\rm in}+D_{\rm 0U}U_{\rm in}}.
\end{equation}
The modulation efficiency is the ratio between the signal power that is modulated at 4 times frequency of the HWP rotational frequency to the detected power. 
There are two reasons why we defined the modulation efficiency in this way. First, we implicitly assume that the AHWP is going to be rotated continuously and thus, we only pick up the term which is relevant to the 4 times of the rotation frequency already. Secondly, we want to define the efficiency in such that we can make a comparison between the measured degree of polarization and the prediction. This is driven by the fact that it is easy to prepare the fully polarized incident source using a wire grid polarizer. The connection between the lab measured efficiency and the CMB analysis is addressed in T. Matsumura et al.~\cite{tmatsumura}. 
We also use $\phi_{4}$ as the phase of the modulated signal to compare the calculation with the measured data.
For an ideal single HWP, $\Gamma$ becomes Eq.~\ref{mt:birefringent_wo_refl}, we recover the following expression~\cite{tmatsumura}
\begin{equation}
\epsilon=\frac{\sin^{2}\frac{\delta}{2}\sqrt{Q_{\rm in}^{2}+U_{\rm in}^{2}}}{I_{\rm in}+\cos^{2}\frac{\delta}{2}Q_{\rm in}}.
\end{equation}
In the case of $(Q_{\rm in}, U_{\rm in})=(0,1)$, the modulation efficiency is simply proportional to $\sin^{2}\frac{\delta}{2}$.

In our development, we aim for modulation efficiency greater than 0.98 in a frequency range of 34-270~GHz with an incident angles up to 10 degrees for linearly-polarized incident light. 
The value of the modulation efficiency 0.98 is the requirement of LiteBIRD LFT telescope.
The choice of the maximum incident angle is driven by the size of the field-of-view in LiteBIRD LFT telescope~\cite{kashima}.

\section{SAMPLE PREPARATION}
\label{sec:sample_prep}  
\subsection{Design optimization}
The AHWP consists of multiple sapphire plates with optic axes relatively rotated among the plates.
We use sapphire as the birefringent material for the HWP.
Sapphire has superior optical and thermal properties for a HWP: 
about 10\% difference in the refractive indices between the ordinary and the extraordinary rays~\cite{b.r.johnson}, the low loss-tangent at millimeter-wave frequency, and the high thermal conductivity, $10^2$-$10^3$~W/K/m, at a temperature of 4 - 10~K~\cite{frank_pobell}.
According to M. N. Afsar~\cite{afsar}, the variation of the refractive index of sapphire in the frequency range of 60-400 GHz is less than 0.1\% for both ordinary and extraordinary rays.
In this paper, we assume the refractive indexes are constant at 34-270 GHz.

We have conducted the design optimization for the AHWP using a simulation. 
The Mueller matrix of $N$-layer AHWP is described as
\begin{equation}
\Gamma_{\rm AHWP}=\prod_i^N R(-\chi_{i})\Gamma R(\chi_{i}),
\label{mt:mueller_ahwp_wo_refl}
\end{equation}
where $\Gamma$ is the Mueller matrix of a single birefringent plate, $\chi_{i}$ is the orientation of the optic axis with respect to the $x$-axis. 
We assume that all plates of which the AHWP is composed have the same thickness $d_\mathrm{ c } = \frac{1}{2}\frac{c}{\Delta n \nu_\mathrm{ c } }$, and the same refractive indices, where $\nu_\mathrm{ c } $ is the center of the frequency band and $c$ is the speed of light. 
In this process, we did not consider any effects of reflection to save computational time.
After the completion of the optimization, we considered the performance including the reflection using the calculation referring to T. Essinger-Hileman~\cite{essinger}. 
Without any reflections, the Mueller matrix of a birefringent material is give in Eq. \ref{mt:birefringent_wo_refl} with the retardance of $\delta$,
\begin{equation}
\Gamma=\left(
\begin{array}{cccc}
1&0&0&0\\
0&1&0&0\\
0&0&\cos\delta&-\sin\delta\\
0&0&\sin\delta&\cos\delta 
\end{array}
\right).
\label{mt:birefringent_wo_refl}
\end{equation}

The design optimization is carried out using a brute force method.
For the frequency range of 34 - 270~GHz, the center frequency is given as $\nu_c=152$~GHz, and the corresponding thickness is calculated to be 3.14~mm, where
we use $n_\mathrm{o}=3.047$ and $n_\mathrm{e}=3.361$ for the refractive indices at low temperature~\cite{b.r.johnson} and we assume lossless sapphire plates. 
We use the averaged modulation efficiency as the figure-of-merit for this optimization.
We generate random numbers for the relative optic axis angle to find an optimal set of the angles offering the high averaged modulation efficiency in the frequency range. 
We repeated the optimization using various initial starting value to search the wide range of the parameter space in the case of $S_{\rm in}=(1, 0, 1, 0)$. 

\begin{figure}[t]
\begin{center}
\includegraphics[width=\hsize]{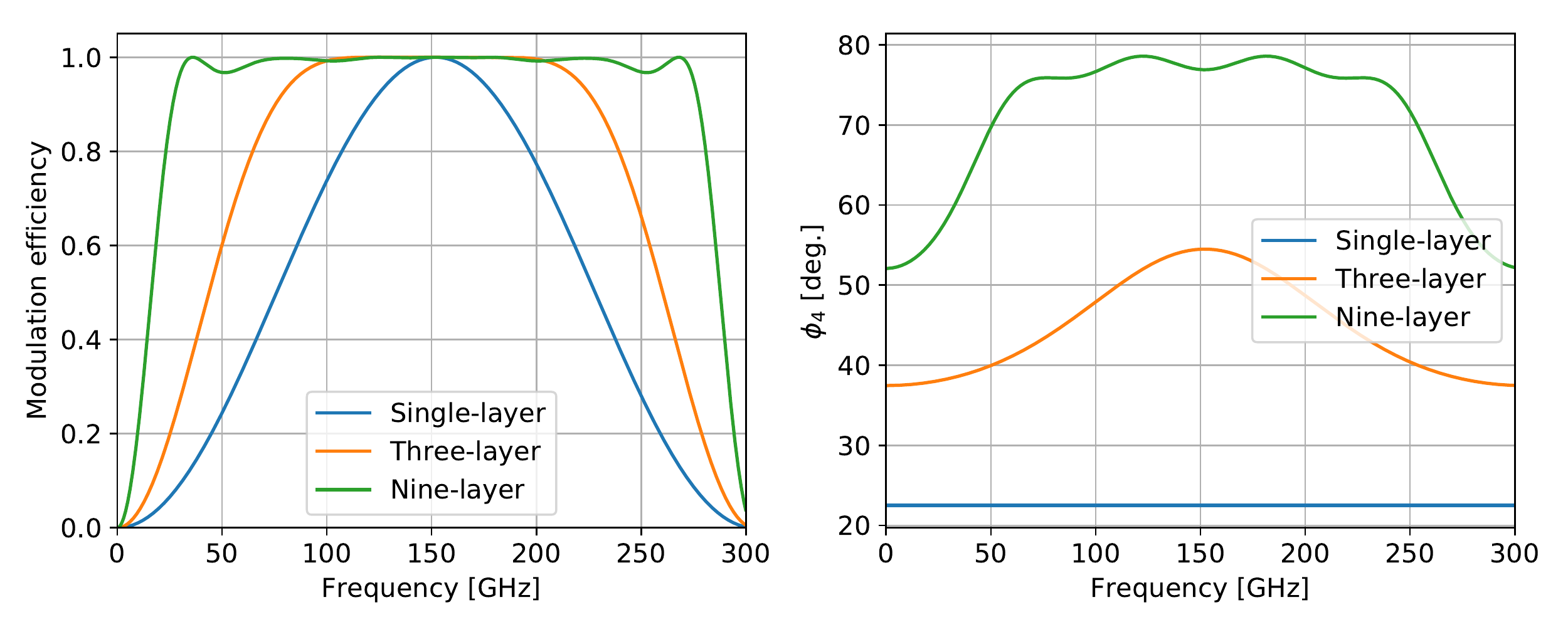}
\end{center}
\caption{\label{fig:design} Calculated modulation efficiency and phase as a function of frequency for the optimized nine-layer AHWP design. For comparison, we also show the calculated result of a single HWP and three-layer AHWP. We calculate for $S_{\rm in}=(1,0,1,0)$.}
\end{figure}

As a result we have concluded that the nine-layer AHWP can cover almost all of the targeted bandwidth.
Figure~\ref{fig:design} shows the modulation efficiency and phase as a function of frequency with the optimized design for a single plate, the three- and nine-layers.
Table~\ref{tab:design} shows the optimized values of the relative angles in nine-layer AHWP. 

\begin{table}[t]
\caption{\label{tab:design} Designed values of the relative angles in the three- and nine-layer AHWP. 
The thickness of each plate is identical. $\chi_{i}$ is the optic axis angle of the $i$-th plate relative to the first layer.
The design of the three-layer AHWP from T. Matsumura at al.~\cite{tmatsumura}.
} 
\begin{center}       
\begin{tabular}{|c|c|c|} 
	\hline
    	number of plates & $d_\mathrm{ c } $ [mm] & $\chi_{i} [^{\circ}]$\\ \hline
    	 3 & 3.14 & 0, 58, 0 \\ \hline
    	 9 & 3.14 & 0, 18.5, 37.5, 73.9, 141.5, 73.9, 37.5, 18.5, 22.7 \\ \hline
\end{tabular}
\end{center}
\end{table}

\subsection{Fabrication}
\begin{figure}[htb]
\begin{center}
\includegraphics[width=0.5\hsize]{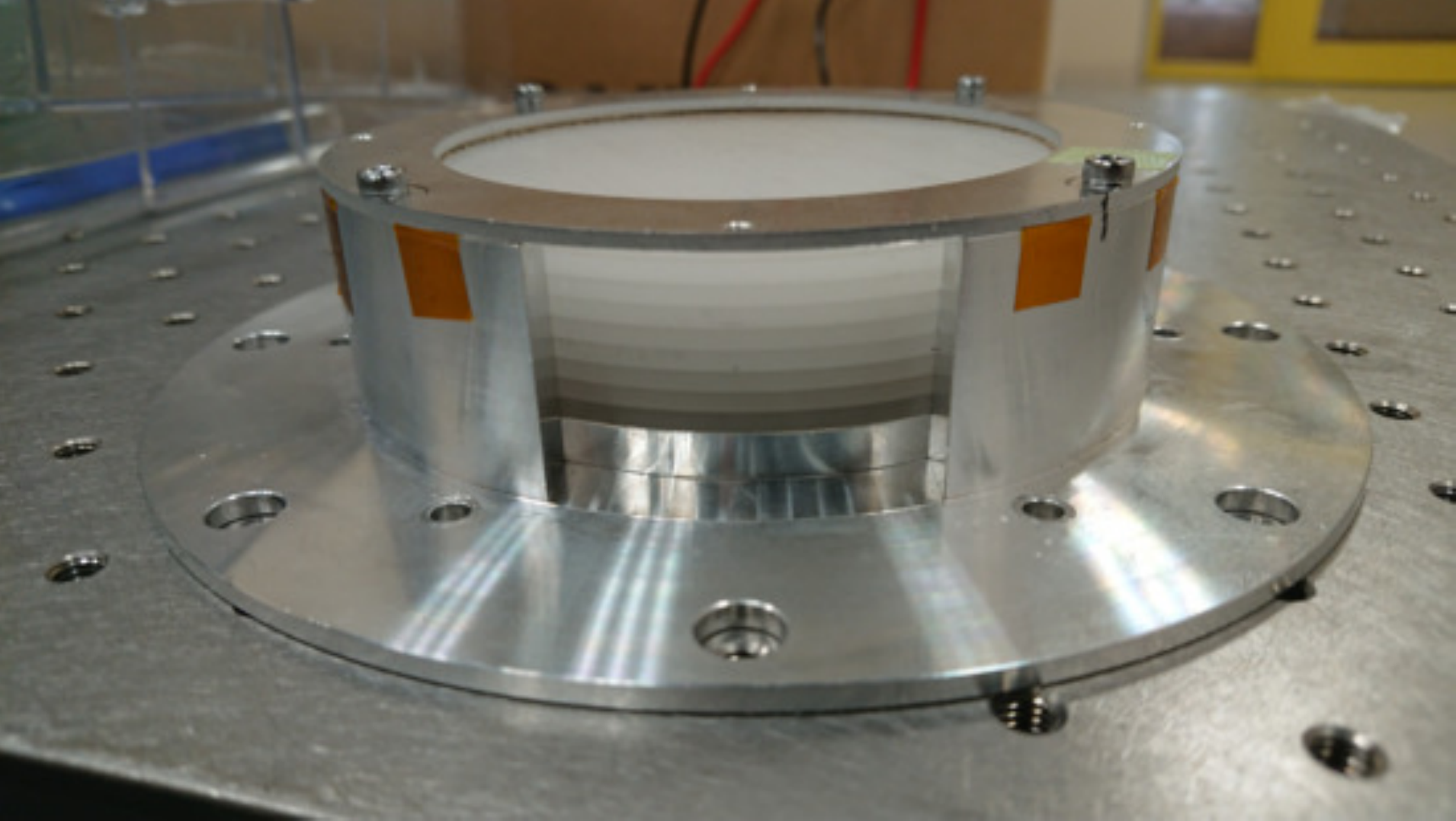}
\end{center}
\caption{The assembled nine-layer AHWP.}
\label{fig:9AHWP_pic}
\end{figure}

We have fabricated the AHWP based on the optimized parameters.
We use a commercially available sapphire sample, with a diameter of 100~mm.
Although the optimized thickness is 3.14~mm for a targeted frequency range, 
we have used sapphire plates with a thickness of 2.53~mm, which was readily available. 
We believe that this slight difference in thickness is not critical for demonstration purposes. 
We measure the thickness of the sapphire along the circumference of the disk, and the variation of the thickness within the sample is found to be less than 8~$\mu$m. 
The surface condition is unpolished. 
We have stacked the plates without glue at the interface of two plates, and fixed them with an aluminum holder as shown in Figure~\ref{fig:9AHWP_pic}.
The anti-reflection coating is not applied on any surfaces.  

To stack the sapphire plates, we use the universal measurement machine (UMM). 
The UMM consists of a microscope and a rotating table. 
Each sapphire plate has an orientation flat (OF) at its side that is in perpendicular to the optic axis and can be used as the reference of the optic axis with the accuracy $3^{\circ}$. 
This accuracy is due to the uncertainty of a dicing capability in a sapphire manufacturing.
Sapphire plates are stacked in the aluminum holder while adjusting with respect to OF to the designed orientation of the optic axis of each plate. 
Figure \ref{fig:9AHWP_pic} shows the assembled nine-layer AHWP.
Sapphire plates are fixed in the holder with pressure applied by an aluminum ring. 
The relative angular uncertainty between the OF of the plates is less than 10~arcmin. 
The implication of this uncertainty is addresssed in Section 6.1.1.

After assembly, we evaluate the presence of air gaps between the layers. 
We investigate the thickness of the air gap by inserting several thin stainless steel plates with varying thickness ($\geq$50~$\mu$m), between layers of the AHWP.
We identify that the air gap between the first and second layer is around 50~$\mu$m, and that air gaps between other layers are less than 50~$\mu$m. 
We will account for the presence of the gap in the analysis.

\section{EXPERIMENT}
\label{sec:exp}  
\subsection{Experimental setup}
\begin{figure}[h]
\begin{center}
\includegraphics[width=0.8\hsize]{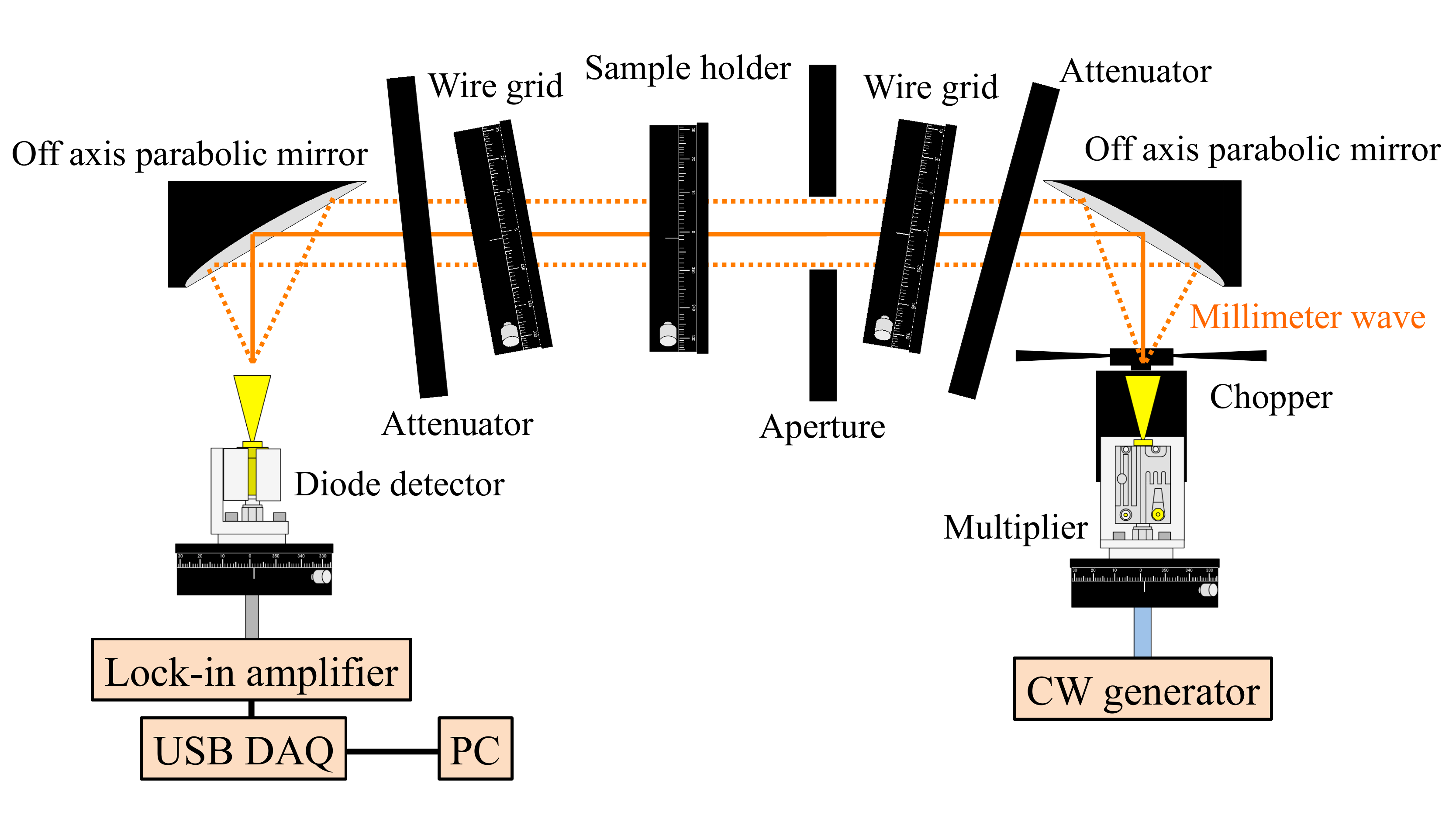}
\caption{A sketch of the measurement system. The millimeter waves propagate along the orange lines from the right hand to the left hand side of the figure. 
The aperture size is approx. 70~mm in diameter.}
\label{fig:setup}
\end{center}
\end{figure}

Figure~\ref{fig:setup} shows the setup for measuring the modulation efficiency and the transmittance in the frequency ranges of 33-140~GHz and 150-260~GHz. 
The millimeter waves are generated by a Continuous Wave (CW) generator and six different active multipliers. 
The CW generator can generate microwaves up to 20~GHz. 
Active multipliers up-convert the frequency of the signal from the CW generator.  
The multiple factors and bandwidths of individual active multipliers are $\times$4 (33-50~GHz), $\times$4 (50-75~GHz), $\times$6 (75-110~GHz), $\times$8 (90-140~GHz), $\times$12 (150-220~GHz) and $\times$24 (210-260~GHz).  
We pair the active multiplier with a diode detector for the measurements in each band. 
Two feedhorns for the source (the active multiplier) and for the detector are placed at the foci of the off-axis parabolic mirrors. 
The millimeter waves emitted from the source horn are collimated by the first mirror.
The plane waves propagate through the first attenuator, the first wire grid,
the 70~mm diameter aperture,
the sample to measure, the second attenuator, and the second wire grid.
The plane waves are focused by the second mirror to be fed to the detector horn.
The millimeter wave from the source is linearly polarized. 
The detector used is a single polarization-sensitive detector. 
We use two free-standing wire grids to define the polarization angle of incident light to the sample.
The transmittance of the wire grid for the light having the electric field perpendicular (parallel) to the wire orientation is 0.99 (0.01) in our measurement frequency range. 
The signal is modulated by an optical chopper with a frequency of 80~Hz to be amplified by a lock-in amplifier. 
The detector outputs the detected power in voltage.
We set the transmission axis of the two wire grids in parallel.
All the measurements were performed at room temperature.
An optical measurement of AHWP at cryogenic temperatures is also in progress and it is beyond the scope of this paper.

\subsection{Measurement of refractive index}
\par To predict the modulation efficiency, we need to know the two refractive indices of an A-cut sapphire plate at the room temperature.
We obtain the indices by measuring transmittance, i.e. Fabry-P\'{e}rot interference, using the same setup in Figure~\ref{fig:setup}.
When the plane of polarization of the incident wave and the detector-sensitive direction are parallel to the optic axis, transmittance $T$ and the complex refractive index $\tilde{n}$ of an extraordinary (ordinary) ray are simply related by Eq.~\ref{eq:trans} for the case of the normal incidence~\cite{hecht}. 
Since the loss tangent of sapphire is small enough to use the approximation, $\tilde{n}$ is expressed using the refractive index $n$ and the loss tangent $\tan\delta$ in Eq.~\ref{eq:c_index}. 
The variable $d$ is the thickness of the sapphire plate and $k_{0}$ is the wave number in the vacuum,
\begin{eqnarray}
\label{eq:trans}  
T(\tilde{n})&=&\left| \frac{2\tilde{n}}{2\tilde{n} \cos k_{0}\tilde{n}d + i(\tilde{n}^{2}+1) \sin k_{0}\tilde{n}d} \right|^{2}, \\
\label{eq:c_index}
\tilde{n}&\simeq&n \left(1-\frac{i}{2}\tan\delta \right).
\end{eqnarray}
We measure the indices of one sapphire plate, and assume the same indices for the rest, which is valid because all the samples are originated from the same batch.
We set the sapphire plate to the sample holder in Figure \ref{fig:setup}.
We then measure output voltages of the lock-in amplifier at 33 to 140~GHz and 150 to 260~GHz in $\sim$ 1~GHz interval. 
After removing the A-cut sapphire plate, we remeasure output voltages of the lock-in amplifier for the same frequency range with the same frequency step.  
The source is a coherent source, and thus the measurement system is susceptible to the effect of the standing wave. 
We mitigate the effect of a standing wave by moving the detector along the optical path by $\lambda/4$ at each frequency.
The spectral shape of transmittance for polarization parallel and perpendicular to the optic axis are computed by taking a ratio of the acquired data between the sapphire plate case and the air case.

\subsection{Measurement of modulation efficiency}
\par  The nine-layer AHWP is mounted on a sample holder which can be automatically rotated by a stepping motor. 
The sample holder continuously rotates around the optical axis in this system with a revolution rate of $\omega_{\rm hwp}=2\pi f_{\rm hwp}$, where $f_{\rm hwp}$ is about 0.02~Hz.
The frequency of the electromagnetic source is swept during the rotation.
We measure the modulated signal as the output voltage of the lock-in amplifier.
The measured frequency range is 33 to 140~GHz and 150 to 260~GHz in $\sim$ 1~GHz interval. 
The measurement time at each frequency is 60 seconds, during which the AHWP rotates about 360 degrees.
The sampling rate of the demodulated signal from the lock-in amplifier is 100~Hz.
For each frequency, we fit the acquired data using Eq. \ref{eq:fit_TOD}. 
\begin{equation}
I(t, \nu) = a_{0}(\nu)+\sum^{8}_{m=1}a_{m}(\nu)\cos{(m \omega_{\rm hwp}t+ m\phi_{m}(\nu))}.
\label{eq:fit_TOD}
\end{equation}
While acquiring data at all frequencies, the AHWP keeps on continuously rotating.
The initial value of $\phi_{n}$ is different for each frequency.
This offset is recorded and subtracted for each $\phi_{n}$ at given frequyency. 
The $m=4$ component is the modulation signal of the AHWP. 
The $m=2$ component appears due to the differential indices which results the frequency dependent transmittance, reflectance, and emissivity between for the ordinary and extraordinary rays of the HWP. 
The other components are included to capture all the features even though we do not expect the odd $m$ components within the framework of the formalism in this paper.
We will address more on this point in the section \ref{sec:discuss}. 
The modulation efficiency and the phase are obtained as $a_{4}/a_{0}$ and $\phi_{4}$, respectively. 
The initial value of $\phi_{4}$ is determined by the initial rotation angle of the sample holder.
We have repeated the measurements for incident angles reletive to the AHWP of 0 and $\pm10^{\circ}$ for $p$- and $s$-polarization, which corresponds to the field-of-view of LiteBIRD and the CMB telescope that observes small angle scales.

In many CMB experiments, the intensity of the observation signal is integrated by the detector with a specific bandwidth.
Therefore, we introduce the band average modulation efficiency to evaluate the integrated modulation signal. 
For the evaluation, we use the frequency bands centered at 40, 50, 60, 68, 78, 89, 100, 119, 140, 166, 195, and 235~GHz with the bandwidth of about 30\%, that are covered by the LiteBIRD LFT.  
We normalize $I(t, \nu)$ by $a_{0}$ for each frequency, and integrate this normalized modulated signals as
\begin{equation}
\int_{\nu_{i}}^{\nu_{f}} \frac{I(t,\nu)}{a_{0}(\nu)} d\nu = \sum^{\nu_{f}}_{\nu=\nu_{i}}\frac{I(t,\nu)}{a_{0}(\nu)} = A_{0}+\sum^{8}_{m=1}A_{m}\cos{(m \omega_{\rm hwp}t+ m\phi_{m})},
\label{eq:integ_TOD}
\end{equation}
\noindent where $\nu_{i}$ and $\nu_{f}$ are the lower and higher boundary in each frequency band, respectively.  
We define $A_{4}/A_{0}$ as the band-averaged modulation efficiency.

\section{RESULTS}
\label{sec:result}  
\subsection{Measurement of refractive index}

\begin{figure}[h]
\begin{center}
\includegraphics[width=\hsize]{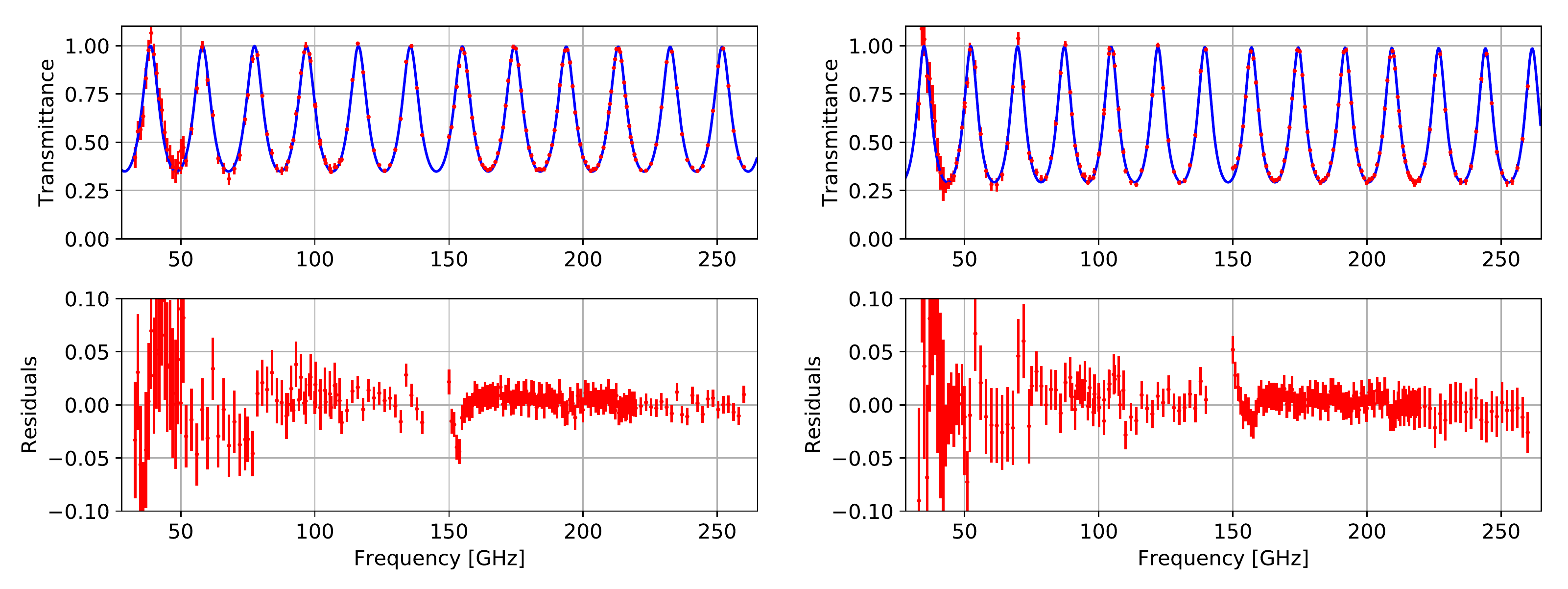}
\end{center}
\caption{\label{fig:trans} The transmittance for the ordinary ray (left) and the extraordinary ray (right) for a A-cut sapphire plate. The top plot shows the measurement (the red dots with error-bars) and fitted (the blue solid line) results. The bottom plot shows the residuals of the fitting. }

\end{figure}

From the fitting result for the frequency-dependent transmittance of the sapphire plate using Eq. \ref{eq:trans} (Figure~\ref{fig:trans}), we obtain the values of the refractive index and the loss tangent at room temperature for the A-cut sapphire plate, which are summarized in Table \ref{tab:no_and_ne}. 
In Figure~\ref{fig:trans}, the difference between the fit and the measured data becomes larger at the lower frequency. 
We think this is caused by the stability of the source or the effect of the diffraction at the aperture, which prevents the full cancellation of the standing effect. (Figure~\ref{fig:setup}).

\begin{table}[ht]
\caption{Fitted result to the refractive index and loss tangent for an A-cut sapphire plate at the room temperature. } 
\label{tab:no_and_ne}
\begin{center}       
\begin{tabular}{|c|c|c|c|c|c|} 
\hline
 \multicolumn{2}{|c|}{Ordinary ray} &  \multicolumn{2}{c|}{Extraordinary ray}  \\
\hline
Refractive index & Loss tangent~($\times 10^{-4}$) & Refractive index & Loss tangent~($\times 10^{-4}$) \\
\hline
 $ 3.059 \pm 0.002 $ & $ 0.9  \pm  0.3 $ & $ 3.397 \pm 0.003 $& $ 1.6  \pm  0.5$ \\
\hline
\end{tabular}
\end{center}
\end{table} 

\subsection{Measurement of modulation efficiency}

\par
\begin{figure}[ht]
\begin{center}
\includegraphics[width=0.6\hsize]{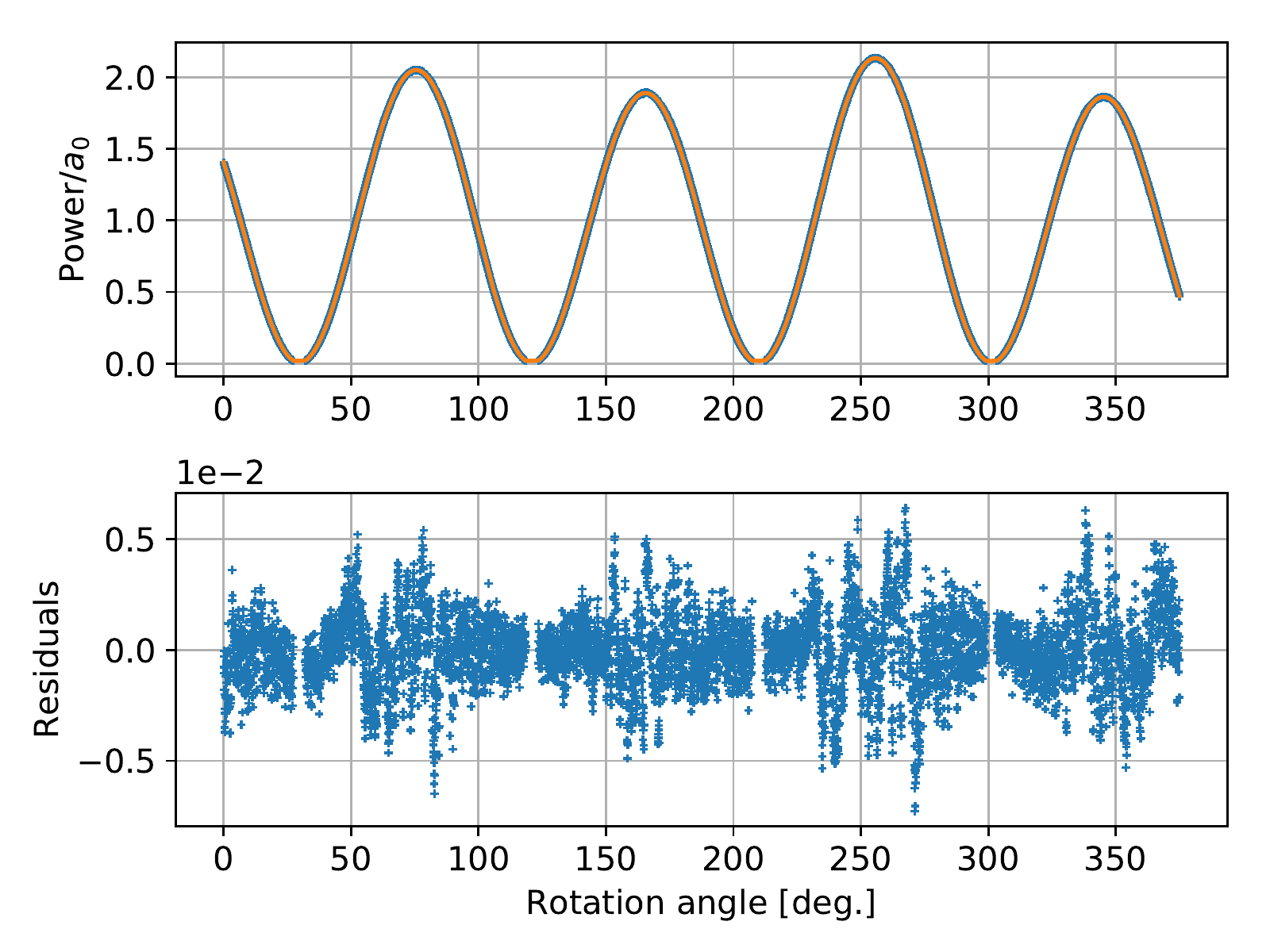}
\caption{The output voltages of the lock-in amplifier as a function of the HWP rotation angle at 150 GHz for the normal incident angle. The output voltage is proportional to the millimeter wave power injected into the detector and normalized by $a_{0}$. The top plot shows the measurement (the blue crosses) and fitted  (the orange solid line) results. The bottom plot shows the residuals of fitting.}
\label{fig:signal}
\end{center}
\end{figure}

\begin{figure}[h]
\begin{center}
\includegraphics[width=\hsize]{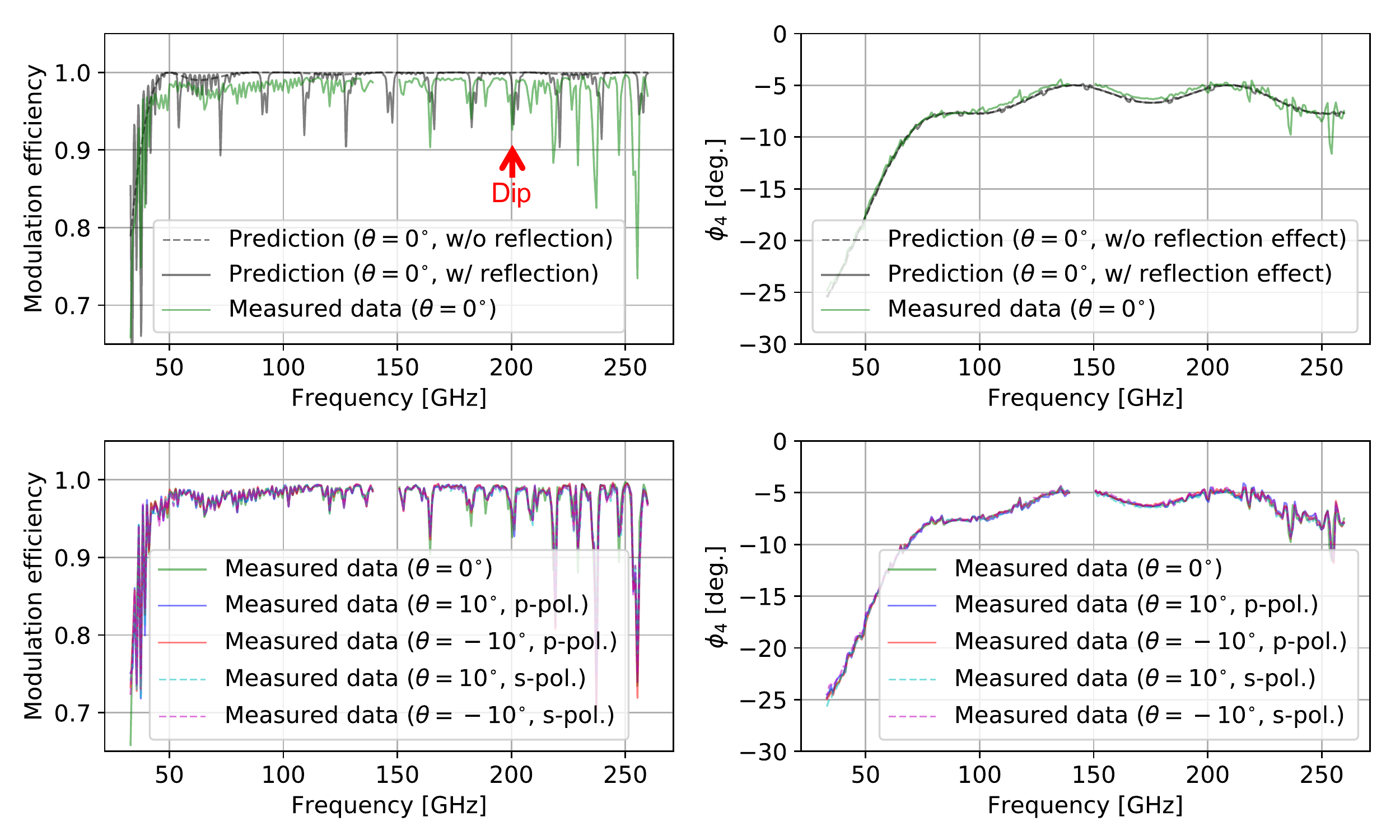}
\caption{\label{fig:overview} The modulation efficiency and the phase from 33 to 260~GHz are plotted in $\sim$ 1~GHz interval, where $\theta$ is the incident angle of the millimeter waves for the AHWP. The predictions are plotted in in 0.2~GHz interval. The top side panels show the comparison of the measured data and the prediction for normal incidence. The bottom side panels show the comparison of normal and oblique incidence.}
\end{center}
\end{figure}

 Figure \ref{fig:signal} shows one example for the output voltages of the lock-in amplifier and fitted result using Eq. \ref{eq:fit_TOD} as a function of the rotation angle at 150~GHz.
The output voltage is normalized by the DC $m=0$ component.
Since the lock-in amplifier can not output a negative voltage, the output signal is inverted when the signal becomes negative due to noise or offset.
This causes the amplitude of the modulation signal to be reduced and causes the modulation efficiency to be underestimated.
Therefore, the part of the modulated signal close to zero is removed. 
For all frequencies and incident angles, we confirm that the residual is less than 3\% (in RMS) of the DC $m=0$ component. 
Figure \ref{fig:overview} shows the frequency dependence of the modulation efficiency and the phase for each incident angle $\theta$ with $p$- and $s$-polarization.
The prediction for the normal incidence takes into account the reflections at plates  and does not consider the air gaps between the plates.
In the prediction, we use the values of $n_\mathrm{o}$ and $n_\mathrm{e}$ at the room temperature in Table \ref{tab:no_and_ne}. 
We can see two features in Figure~\ref{fig:overview}: the sharp dips that appear at about every 18 GHz, and the fast oscillatory features that fluctuate quickly and with a small amplitude.
The dips originate from Fabry-P\'{e}rot interference within each plate that composes the AHWP.
By contrast, the oscillatory feature is from the reflection at the boundaries between the first/last plate and air. 
\par Table~\ref{tab:modeff} shows the measured band-averaged modulation efficiency and Table \ref{tab:phasediff} shows the maximum difference of the phase variation within a bandwidth, which is defined as the LiteBIRD frequency band. 
About the error source of the modulation efficiency is discussed in section 6. 

\begin{table}[ht]
\caption{The measured band-averaged modulation efficiency within the bandwidth for the nine-layer AHWP at each incident angle. } 
\label{tab:modeff}
\begin{center}       
\begin{tabular}{|c|c|c|c|c|c|c|c|} 
	\hline
	\multicolumn{2}{|c|}{} & \multicolumn{5}{c|}{band-averaged modulation efficiency} \\
	\cline{3-7}
    \multicolumn{2}{|c|}{} & $\theta=0^{\circ}$ & $\theta=10^{\circ}$& $\theta=10^{\circ}$ & $\theta=-10^{\circ}$ & $\theta=-10^{\circ}$   \\ \cline{1-2}
  	    	band [GHz] & bandwidth [\%] &  & (p-pol.) & (s-pol.) & (p-pol.) & (s-pol.)   \\   \hline
	 40 & 30 & 0.902 & 0.895& 0.892 & 0.897 & 0.900   \\  \hline
         50 & 30  & 0.961  &  0.960 & 0.959  & 0.961  & 0.960   \\  \hline
         60 & 23  & 0.971  &  0.970  & 0.970  &  0.970  & 0.971   \\  \hline
         68 & 23  & 0.969  &  0.969  &  0.969  & 0.969  & 0.970   \\  \hline
         78 & 23 & 0.976  &  0.975 &  0.976  & 0.976  & 0.977   \\  \hline
         89 & 23 & 0.981  &  0.981  &  0.981  & 0.982  & 0.982   \\  \hline
         100 & 23 & 0.985  &  0.985  &  0.986  & 0.985  & 0.986   \\  \hline
         119 & 30 & 0.984  &  0.983  &  0.983  & 0.983  & 0.984   \\  \hline
         140 & 30 & 0.984  &  0.984  &  0.984  & 0.984  & 0.984  \\ \hline
         166 & 30 & 0.983  &  0.984  &  0.983  & 0.984  & 0.983   \\ \hline
         195 & 30 & 0.979  &  0.979  &  0.978  & 0.979  & 0.979  \\ \hline
         235 & 30 & 0.959  &  0.955   &  0.955  & 0.954 & 0.954  \\ \hline
\end{tabular}
\end{center}
\end{table} 

\begin{table}[ht]
\caption{The maximum difference of the phase variation within the bandwidth for the nine-layer AHWP at each incident angle. } 
\label{tab:phasediff}
\begin{center}       
\begin{tabular}{|c|c|c|c|c|c|c|} 
	\hline
	\multicolumn{2}{|c|}{} & \multicolumn{5}{c|}{$\Delta\phi_{4}$} \\
	\cline{3-7}
    \multicolumn{2}{|c|}{} & $\theta=0^{\circ}$ & $\theta=10^{\circ}$& $\theta=10^{\circ}$ & $\theta=-10^{\circ}$ & $\theta=-10^{\circ}$   \\ \cline{1-2}
  	    	band [GHz] & bandwidth [\%] &  & (p-pol.) & (s-pol.) & (p-pol.) & (s-pol.)   \\   \hline
	 40 & 30 & $4.86^{\circ}$  &  $5.11^{\circ}$  &  $5.34^{\circ}$  & $5.05^{\circ}$  & $4.77^{\circ}$   \\  \hline
         50 & 30 & $6.86^{\circ}$  &  $7.25^{\circ}$  &  $6.78^{\circ}$  & $7.10^{\circ}$  & $6.64^{\circ}$   \\  \hline
         60 & 23 & $5.20^{\circ}$  &  $5.41^{\circ}$  &  $5.08^{\circ}$  & $5.56^{\circ}$  & $5.01^{\circ}$   \\  \hline
         68 & 23 & $3.90^{\circ}$  &  $4.17^{\circ}$  &  $3.82^{\circ}$  & $3.97^{\circ}$  & $3.86^{\circ}$   \\  \hline
         78 & 23 & $2.52^{\circ}$  &  $2.86^{\circ}$  & $2.74^{\circ}$  & $2.65^{\circ}$  & $2.47^{\circ}$  \\  \hline
         89 & 23 & $0.82^{\circ}$  &  $0.79^{\circ}$  & $0.68^{\circ}$  &  $0.69^{\circ}$  & $0.74^{\circ}$   \\  \hline
         100 & 23 & $0.95^{\circ}$  &  $0.94^{\circ}$  &  $0.83^{\circ}$  & $0.89^{\circ}$ &  $1.05^{\circ}$   \\  \hline
         119 & 30 & $3.15^{\circ}$  &  $3.05^{\circ}$  & $3.25^{\circ}$ & $3.13^{\circ}$  & $3.22^{\circ}$   \\  \hline
         140 & 30 & $1.83^{\circ}$  &  $1.83^{\circ}$  &  $1.97^{\circ}$ & $1.86^{\circ}$   & $1.82^{\circ}$  \\ \hline
         166 & 30 & $1.49^{\circ}$  &  $1.51^{\circ}$  &  $1.41^{\circ}$ & $1.51^{\circ}$   & $1.74^{\circ}$   \\ \hline
         195 & 30 & $2.08^{\circ}$  &  $2.31^{\circ}$ & $1.74^{\circ}$ & $1.98^{\circ}$  & $2.00^{\circ}$  \\ \hline
         235 & 30 & $7.12^{\circ}$  &  $7.65^{\circ}$  &  $6.63^{\circ}$& $7.37^{\circ}$   & $6.84^{\circ}$  \\ \hline
\end{tabular}
\end{center}
\end{table} 

\section{DISCUSSION}
\label{sec:discuss}  
\subsection{Sources of error}
Figure~\ref{fig:overview} shows that the prediction and the measurement data are in good agreement for both modulation efficiency and phase. 
Nevertheless, we identify some discrepancies between the measurement results and the predictions.
Here, we discuss possible sources of the the discrepancies between the measurement results and the predictions.

\subsubsection{Relative angular uncertainties}
\begin{figure}[ht]
\begin{center}
\includegraphics[width=\hsize]{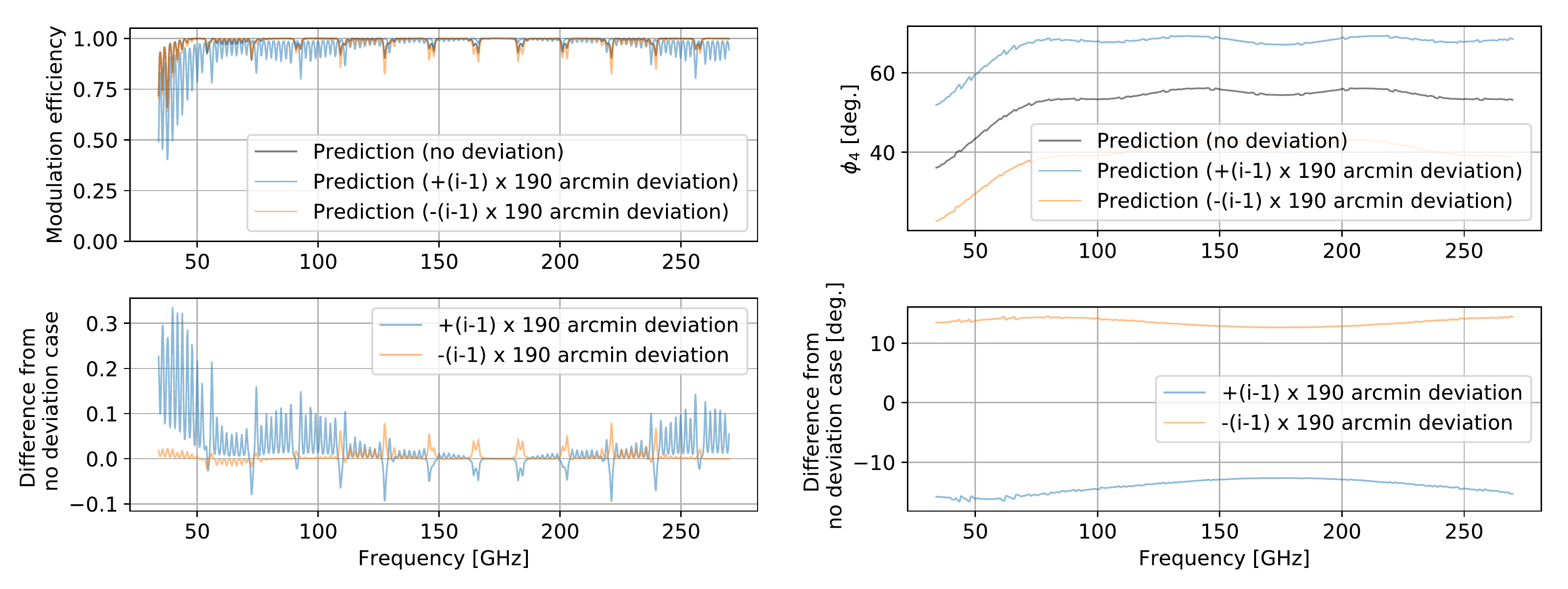}
\end{center}
\caption{Calculated results of the modulation efficiency and the phase for the nine-layer AHWP with and without the angular position shifts.
The bottom plots show the differences from the case with no uncertainty.}
\label{fig:angle_err_effect}
\end{figure}

In the fabrication of the nine-layer stacked AHWP using the UMM,
we find the relative angle position error of OF is less than 10~arcmin.
The OF can be used as the reference of the optic axis with an accuracy of $3^{\circ}$ (180~arcmin).
Therefore the angular position uncertainty of the optic axis of the $i$-th plate is less than $(i-1)\times 190$~arcmin. 
Here we consider a conservative case; all the plates have angular position
shifts of 190~arcmin relative to the former plate in the same direction.
Figure \ref{fig:angle_err_effect} shows the comparison of the modulation efficiency and the phase with and without this angular position shifts.
The bottom plot shows the difference between them.
In this calculation, we ignore the air gaps and fix all the other parameters to
their designed values.
From the comparison result, the differences are found to be less than 0.33 (0.06 in RMS) for the modulation efficiency
and less than $17^{\circ}$ ($14^{\circ}$ in RMS) for the phase.
This $14^{\circ}$ in RMS is obtained in a very conservative way. 
In the fabrication of the AHWP used for the LiteBIRD observation, we expect to use the sapphire plates produced from the same ingot and define each optic axis in higher accuracy by optical measurement nor X-ray without referring OF.
The position determination precision in stacking the plates is expected to be order of 10 arcmin with random variation. 
In addition to this, we can calibrate the modulation efficiency and the phase in the ground facility with an accuracy sufficient to remove the foreground components with a required level. 

\subsubsection{Thickness uncertainty of sapphire plates}
We estimate the thickness uncertainty in individual sapphire plates
to be $\pm$4~$\mu$m from measurements of thickness variation
along the circumference of the disk. 
We therefore calculate the modulation efficiency and the phase by assuming a plate thickness of 2.534~mm and 2.526~mm and compare it with the calculated result of the plate thickness of 2.530~mm.
We again ignore the air gaps and set all the other parameters to designed values.
From the comparison, the uncertainties of the modulation efficiency and the phase are estimated to be less than 0.05 (0.007 in RMS) and  $0.4^{\circ}$ ($0.06^{\circ}$ in RMS), respectively.

Note that the finite thickness itself of the AHWP can result a potential systematic effect.
In this paper, the total thickness of the AHWP is about 23~mm. 
When such a large thickness AHWP is employed with a converging or a diverging optical system, the focus position might be affected.
In case of LiteBIRD, the AHWP is placed as a first optical element, and thus the incident radiation is in parallel. 
As a result, this particular issue is not a relevant unless we further look into the higher order effect.
This is beyond the AHWP design issue but rather a system wide design issue, which depends on what optical system to use with the AHWP. 
Therefore, we do not address further in this paper.

\begin{figure}[ht]
\begin{center}
\includegraphics[width=\hsize]{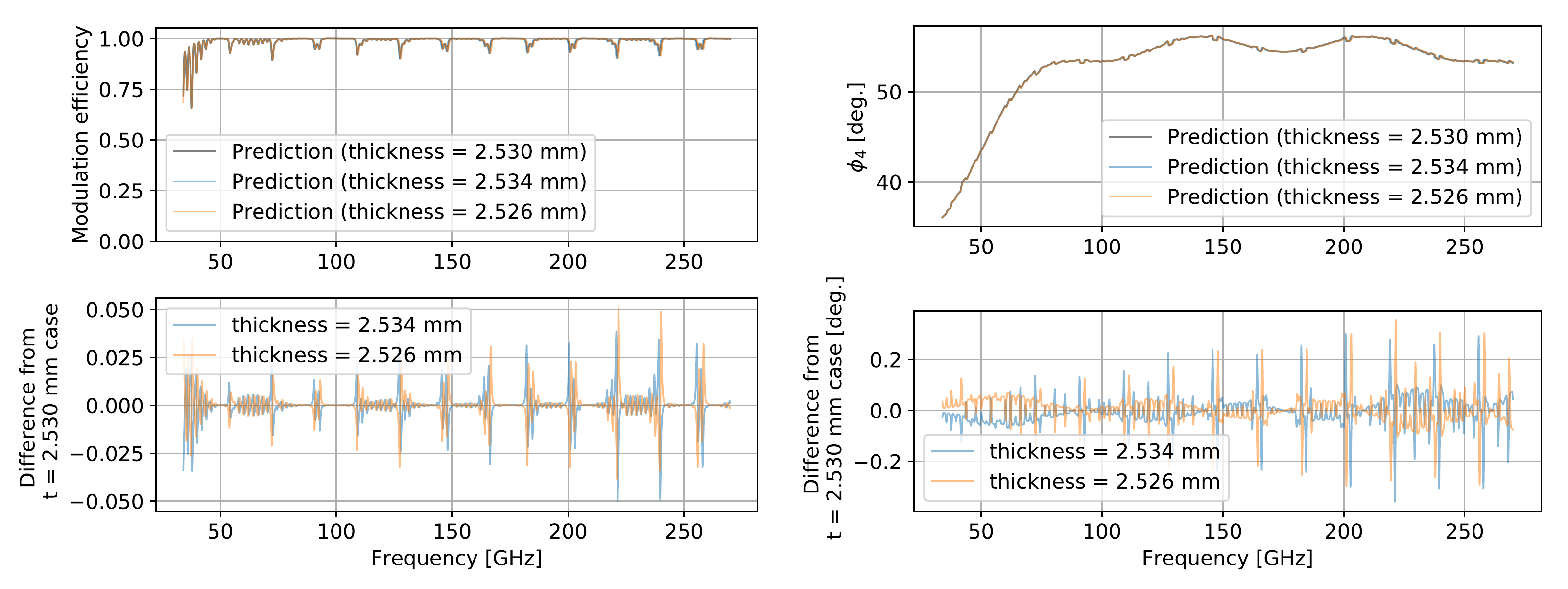}
\end{center}
\caption{Calculated results of the modulation efficiency and the phase of the nine-layer AHWP for the plate thicknesses of 2.530~mm, 2.534~mm and 2.526~mm. The bottom plot shows the differences of those values from the ones with 2.530~mm.}
\label{fig:thickness_err_effect}
\end{figure}

\subsubsection{Refractive indices uncertainty}
From Table \ref{tab:no_and_ne}, we estimate the difference between the two refractive indices of the A-cut sapphire to be $\Delta n = 0.338 \pm 0.005$.
We compare the calculation results that $\Delta n$ set to 0.343,  0.333 and 0.338.
Here we ignore the air gaps and set all the other parameters to the designed values.
With the comparison, we find the uncertainties in the modulation efficiency and phase to be 0.03 (0.003 in RMS) and $0.4^{\circ}$ ($0.2^{\circ}$ in RMS), respectively.

\begin{figure}[ht]
\begin{center}
\includegraphics[width=\hsize]{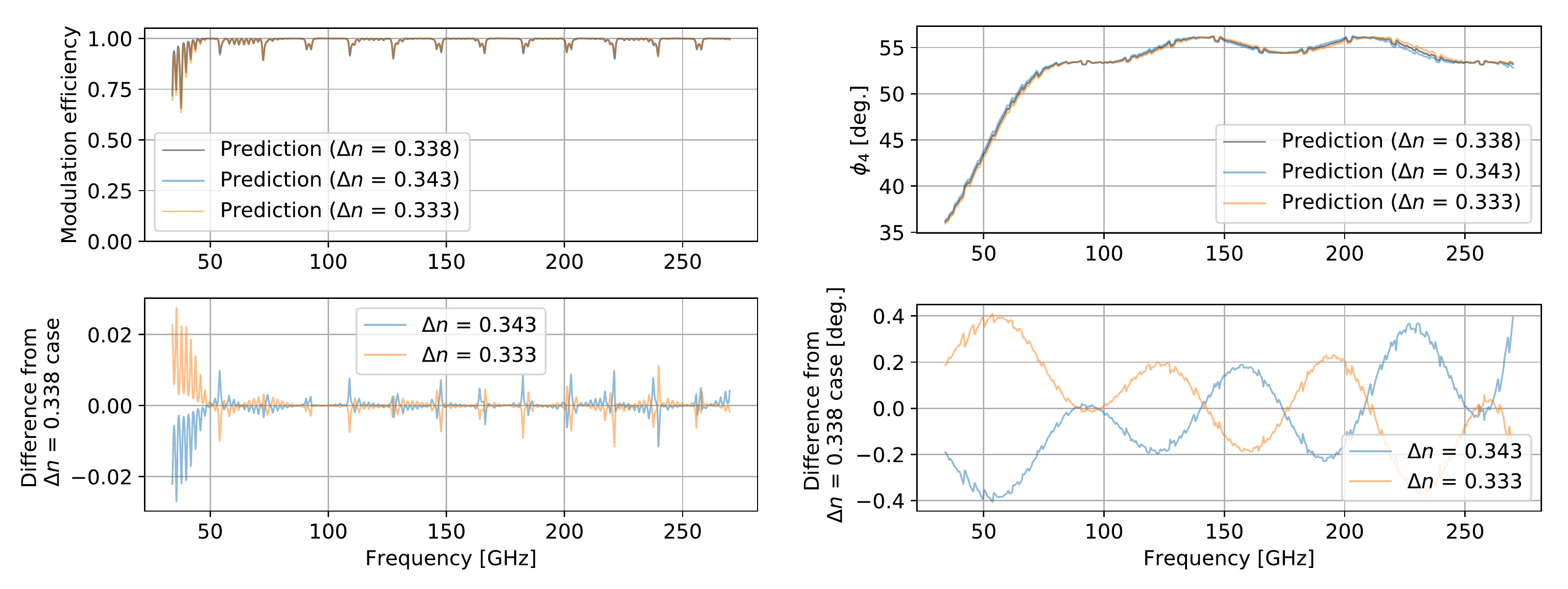}
\end{center}
\caption{Calculated results of the modulation efficiency and the phase of the nine-layer AHWP for the refractive index differences of 0.338, 0.343 and 0.333. 
The bottom plot shows the difference of them from the ones with 0.338.}
\label{fig:index_err_effect}
\end{figure}

\subsubsection{The effect of air gaps}
\begin{figure}[ht]
\begin{center}
\includegraphics[width=\hsize]{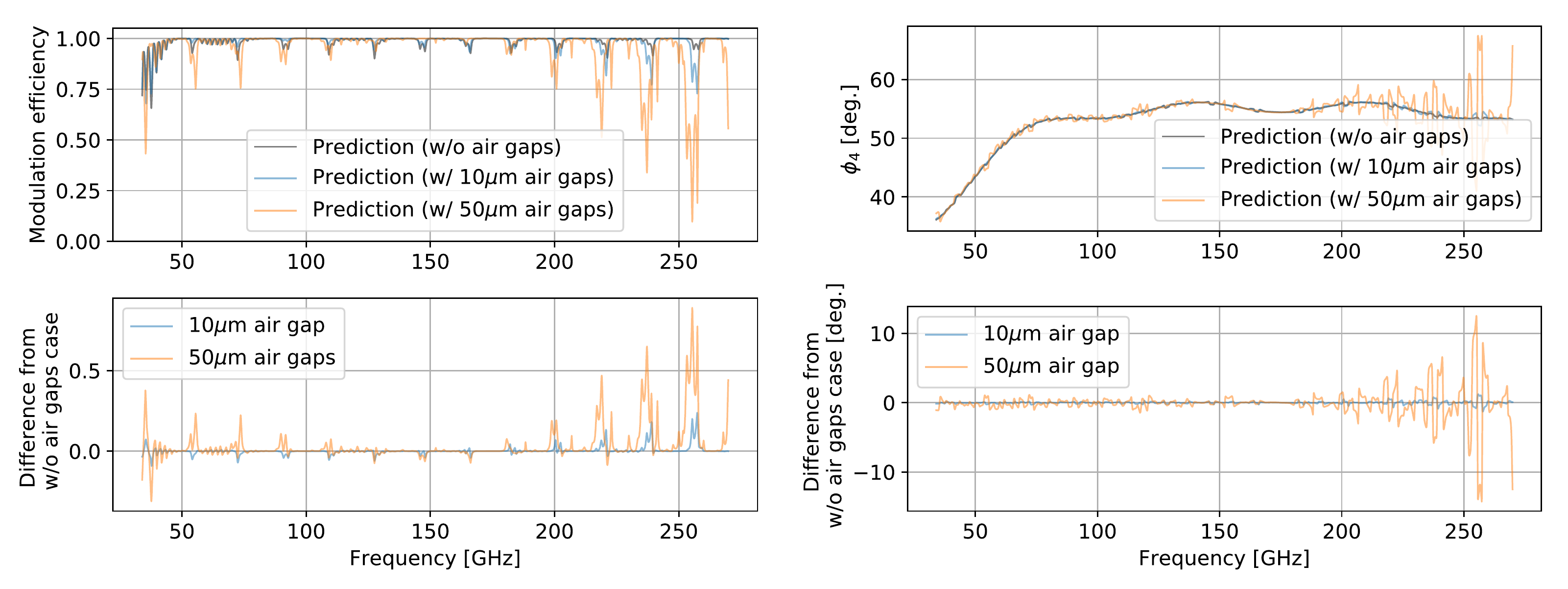}
\end{center}
\caption{Dependence of the modulation efficiency and the phase on the air gaps.
The bottom plot shows the difference of them from the ones without the air gap.}
\label{fig:airgap_effect}
\end{figure}

After the assembly of the AHWP, we identify the air gap between the first and second layer to be around 50~$\mu$m, and air gaps between other plates are found to be less than 50~$\mu$m. 
We compare the modulation efficiency and the phase of the nine-layer AHWP in calculation with and without the 10~$\mu $m and 50~$\mu $m air gaps between all plates. 
Figure \ref{fig:airgap_effect} shows the comparison results.
We find no difference in the modulation efficiency and the phase around 175.3~GHz.
This is because the transmittance of all the sapphire plates is close to unity at this frequency and the air gaps do not contribute to the reflections at the boundaries of the plates.
On the other hand, Figure~\ref{fig:airgap_effect} implies that the dips and oscillatory features seem to depend on the magnitude of the air gap.
This is because those two features originate from the reflections in the AHWP and the air gaps affect those reflections.
For example, on the high frequency side, the depth of the dips and oscillatory feature monotonically increases according to the thickness of the air gap. 
On the low frequency side, where the thickness of the air gap is small, the dip depth is decreased, and where the thickness becomes large, it starts to increase.
This trend is consistent with the difference between the prediction and the measured data in Figure~\ref{fig:overview}.
The air gaps cause the changes in the modulation efficiency and phase to be 0.9 in maximum (0.1 in RMS) and $14^{\circ}$ at maximum ($2^{\circ}$ in RMS), respectively.

We find that the air gaps affect the modulation efficiency and phase around the dips significantly.

\subsubsection{Summary of sources of error}
\begin{figure}[ht]
\begin{center}
\includegraphics[width=\hsize]{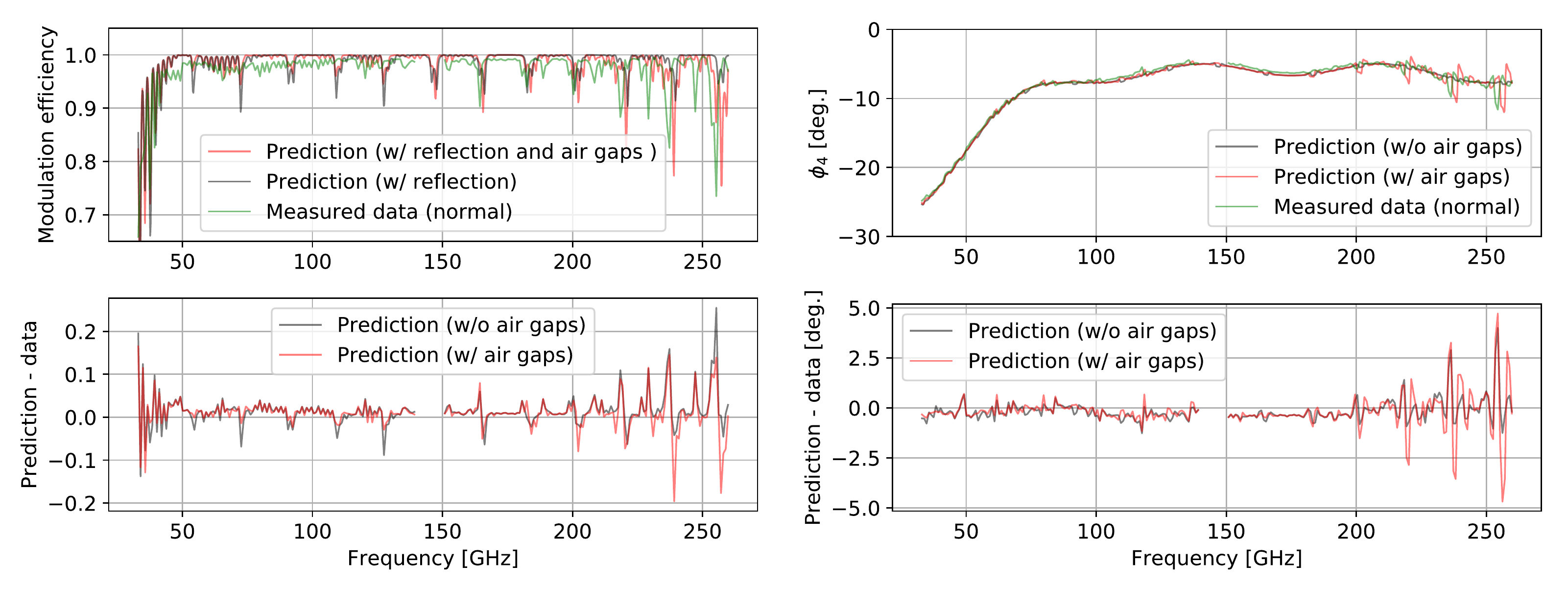}
\end{center}
\caption{\label{fig:overview2} The modulation efficiency (left panels) and the phase (right panels) from 33 to 260~GHz are plotted in $\sim$ 1~GHz interval. The predictions takes into account the air gaps based on the measurements and are plotted in 0.2~GHz interval.}
\end{figure}

From the consideration in this subsection, we find that the air gaps are the largest source of the change in the AHWP performance around the dips.
Figure~\ref{fig:overview2} shows a comparison of the measured data and the prediction that takes into account the largest source, air gaps.
In Figure~\ref{fig:overview2}, the prediction taking the measured air gaps into account reproduces the tendency of the depth of the dip and oscillatory feature better than before.
But the residuals of the prediction and the measured data is not decreased because of the frequency shift of the dips caused by the uncertainties of thickness and refractive indices.

In the prediction, the air gaps are inserted between each plate as a parallel flat plate having a refractive index of 1.
The thickness of the air gap between the first and the second plate is set to 50~$\mu$m and other gaps are set to 8~$\mu$m. 
The thickness of the air gaps other than those between the first and second plates is obtained from the thickness variation along the circumference of the plates.

\subsection{Amplitude of modulated signal for each mode}
\begin{figure}[h]
\begin{center}
\includegraphics[width=0.5\hsize]{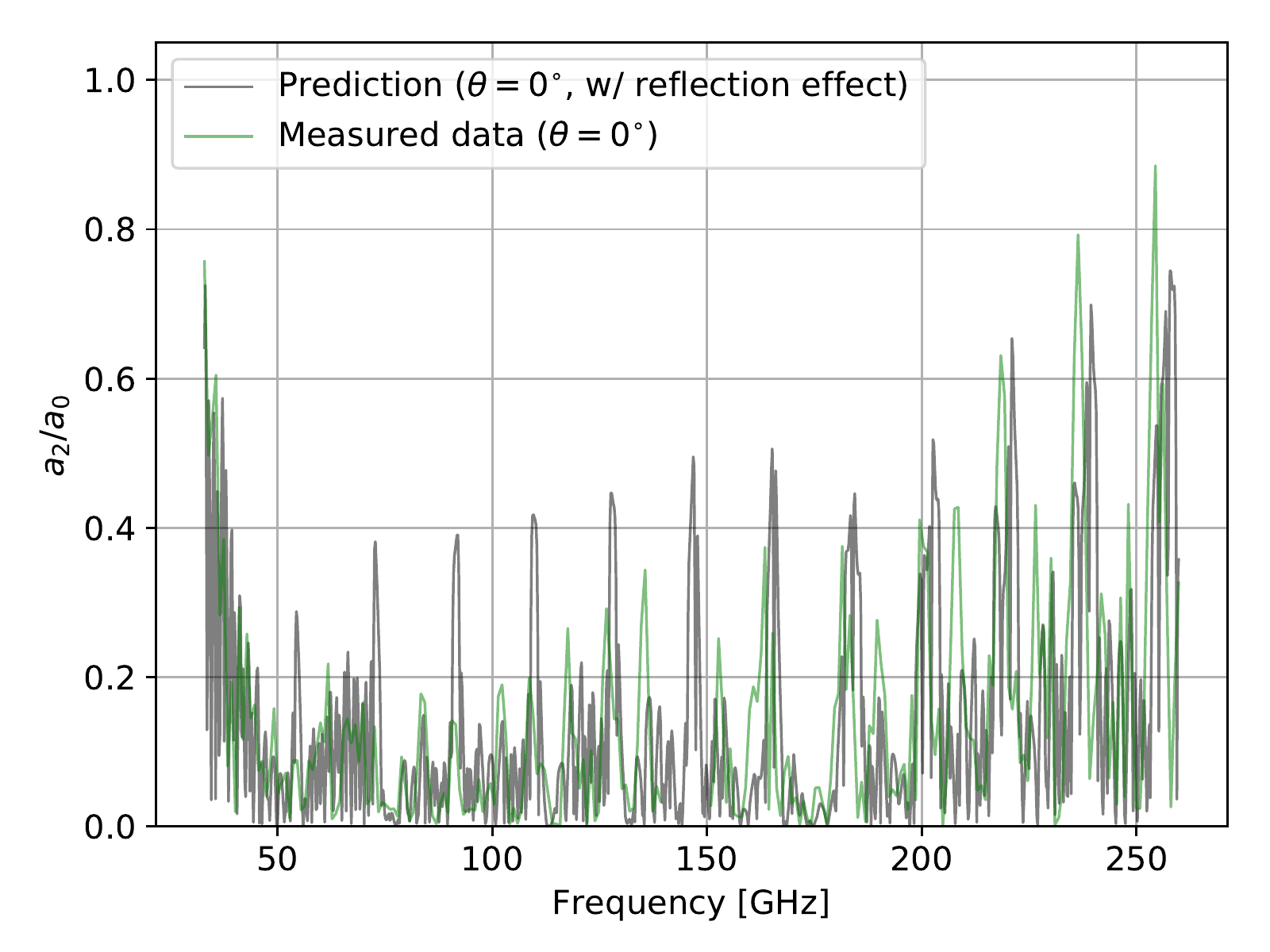}
\caption{\label{fig:overview_2f} The amplitude of $m=2$ mode from 33 to 260~GHz are plotted in $\sim$ 1~GHz interval. The predictions takes into account the air gaps and are plotted in 0.2~GHz interval.}
\end{center}
\end{figure}

\begin{figure}[h]
\begin{center}
\includegraphics[width=\hsize]{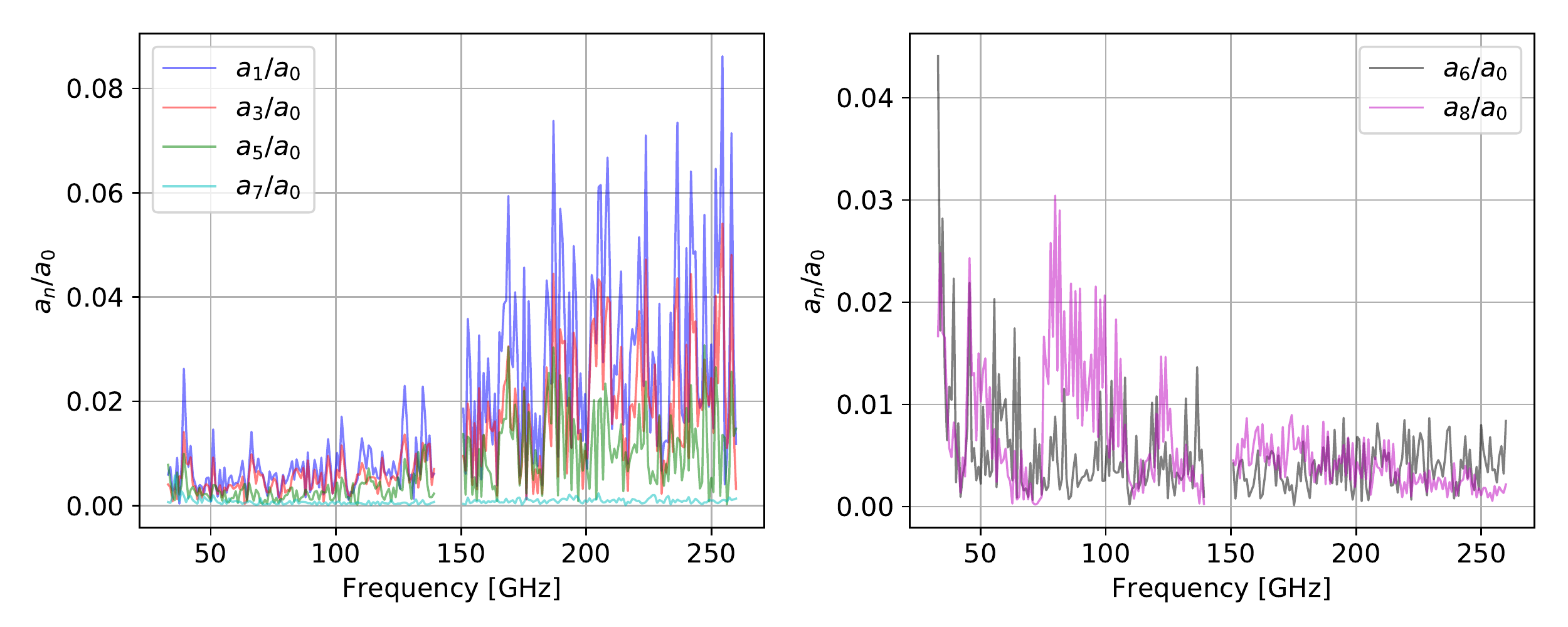}
\caption{\label{fig:overview_nf} The $m=1, 3, 5, 6, 7, 8$ modes from 33 to 260~GHz are plotted in every 0.9~GHz.}
\end{center}
\end{figure}

To obtain the modulation efficiency and the phase, we use Eq. \ref{eq:fit_TOD} for the fitting of the modulated signal.
Here, we discuss the modes other than $m = 4$.
The $m=2$ mode appears due to the difference of the frequency-dependenct transmittance between the two refractive indices in the HWP caused by the absence of the anti-reflection coating. 
Figure \ref{fig:overview_2f} shows a comparison of the measured data and the prediction of the $m=2$ mode which takes into account the air gaps.
While we think the model of the air gap is not complete, we qualitatively recover the agreement between the prediction and the measured data.
The origin of other modes, $m=1, 3, 5, 6, 7, 8$, is not physically motivated within the formalism described in Eq.~\ref{eq:Iout}.
The peak amplitude is generally signal-to-noise above 100 for $m=1, 3, 5, 6, 8$ in the range of above 40~GHz.
Thus, the identified peaks are not due to the noise. 
One of the potential contributors to the peaks at $m \neq 2, 4$ is from the imperfection of the assembly.
Figure~\ref{fig:overview_nf} shows the measured amplitude of the mode at $m=1, 3, 5, 6, 7, 8$.
For the mode $m=1, 3, 5$, we identify the general trend of higher amplitude as it extends to the higher frequency.
This is generally consistent with the effect of the air gap. 
The modes at $m=6, 7, 8$, on the other hand, show the different tendency. The modes $m=6, 8$ originate from the refractive index for an extraordinary ray as a function of an incident angle. When an incident wave enters with an incident angle of $\theta$ and an azimuthal angle of $\phi$ with respect to the optic axis, the refractive index $n_\mathrm{e}^\prime (\theta,\phi)$ is given by\cite{Orfanidis}

\begin{align}
n_\mathrm{e}^\prime (\theta,\phi) = n_\mathrm{e} \sqrt{ 1 - \left( \frac{ 1 } { n_\mathrm{ o }^2 } - \frac{ 1 } { n_\mathrm{ e }^2 } \right) n_1^2 \sin^2 \theta \cos^2 \phi },
\end{align}
where $n_1$ is an index of the ambient space. $n_\mathrm{e}^\prime (\theta,\phi)$ includes higher order cosines of $\phi$. It cannot account for the odd order cosines of $\phi$. This is the reason the modes $m=6, 8$ have the different tendency. However, the reason of the $m=7$ mode tendency remains unclear.

\begin{figure}[h]
\begin{center}
\includegraphics[width=0.6\hsize]{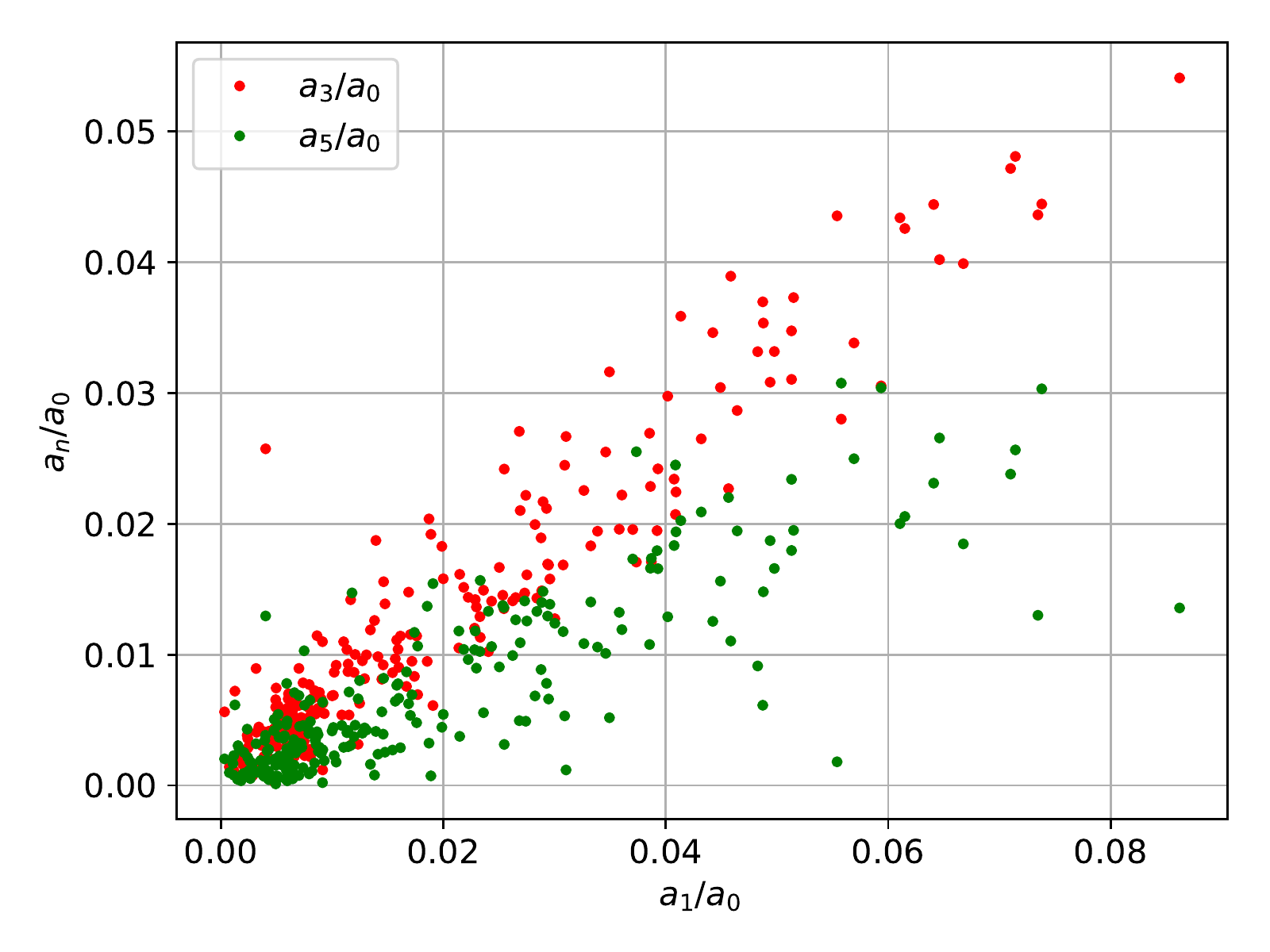}
\caption{\label{fig:comparison_1f_odd} The correlation of $a_{1}/a_{0}$ and $a_{3}/a_{0}$, $a_{5}/a_{0}$ from 33 to 260~GHz. The data is every 0.9~GHz.}
\end{center}
\end{figure}

We also looked at the correlation between $m=1$ and $m=3, 5$ as shown in Figure~\ref{fig:comparison_1f_odd}.
If the majority of the effect is due to the air gap and the air gap has a wedge-like shape, we expect the rotational synchronous $m=1$ mode and potentially higher harmonics. 
Figure~\ref{fig:comparison_1f_odd} shows the positive correlation between the $m=1$ mode and the $m=3, 5$, which supports the idea of the effect of the wedge shape air gap and its harmonics. 
Candidates for other sources of $m=1, 3, 5$ modes are the rotational speed instability and the vibration and wobble in the rotation, and these are expected to be more pronounced at higher frequencies.

The $m\neq4$ modes can be filtered out at the demodulation step.
There is a potential leakage from $m=2$ to higher harmonics as a source of the conversion from unpolarized light to polarized light.
Such a possibility can be addressed particularly when the incident angle is not normal incident to the HWP. 
The result is highly dependent on the performance of the AR coating, which is not accounted for in this paper.
An example of the study can be found in T. Essinger--Hileman et al.~\cite{essinger_spie} and H. Imada et al.~\cite{imada_isstt2018}
Thus, we decide not to explore beyond the identification of the existence of $m\neq4$ mode.

\subsection{Dependency for incident angle}
In Table \ref{tab:modeff} we show the band-averaged modulation efficiencies
with the incident angle of 10 degrees.
The differences of the efficiencies between normal and the 10 degree 
oblique incidence are less than 0.005 (0.01) for the highest (lowest) bands.
For the other bands, the differences are less than 0.001.
The bottom side panels of Figure \ref{fig:overview} show the modulation efficiency and the phase for the two incident angles, suggesting there is no significant difference between them.
\par
\begin{figure}[h]
\begin{center}
\includegraphics[width=\hsize]{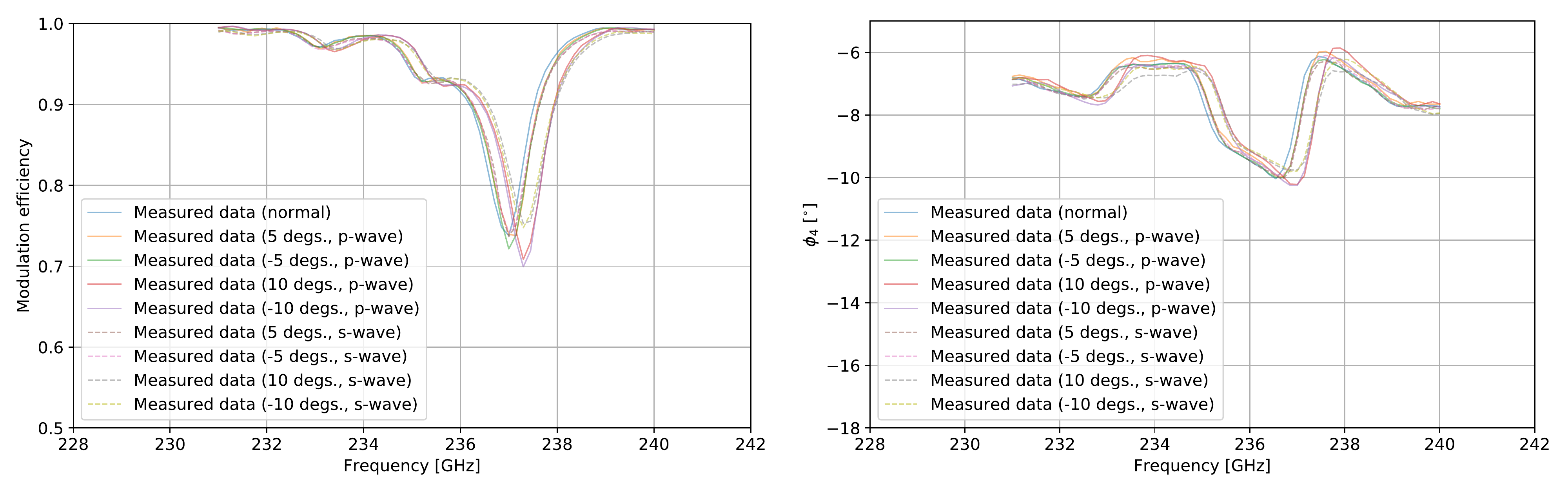}
\end{center}
\caption{Frequency dependence of the modulation efficiency and the phase for 230 to 240~GHz. The data points of the measurement result are plotted for every 0.15~GHz.}
\label{fig:230-240}
\end{figure}

 In order to understand the incident angle dependence more precisely,
we repeat the measurement with a finer frequency step of 0.15~GHz between 230 and 240~GHz,
and with incident angle of $\pm 5^{\circ}$.
Figure~\ref{fig:230-240} shows the incident angle dependencies, and we find that there is a frequency shift of the dip.
In case of the incident angle of  $5^{\circ}$ ($10^{\circ}$),
the refraction angle at the first plate of the AHWP is calculated to be about $1.5^{\circ}$ ($3^{\circ}$) for each ray.
Since the difference in the refractive indices is small at the boundary between the two sapphire plates, the refraction angles within each plate are similar to the first plate.
When $\theta \neq 0$,  frequencies of Fabry-P\'{e}rot interference spectrum scaled by $(\cos \theta)^{-1}$ compared with the normal incidence ($\theta = 0$).
That causes the frequency shift to the higher side.
For the refraction angle of $1.5^{\circ}$ ($3^{\circ}$), the frequency shift is computed to be about 0.08 (0.33)~GHz. 
The estimation of the frequency shift is in good agreement to the shift measured in the modulation efficiency and the phase.  
Therefore, we find the observed frequency shift can be explained by the incident angle dependence.

The dependency for incident angle in AHWP is also related to the potential conversion from the incident unpolarized light to the polarized light. 
This is one of the important effect to be addressed in CMB experiments.
H. Imada et al.~\cite{imada_isstt2018} addressed this effect, and the further study to propagate this effect to quantify the impact to the cosmology is in progress.

\subsection{Further design optimization}
In spite of the fact that the nine layer AHWP becomes broadband as expected, the band-averaged modulation efficiency is lower than 0.98 in some bands.
We find three reasons why the band averaged modulation efficiency is lower than 0.98.
The first two reasons are due to the hardware preparation, and the third one arises from the AHWP design.
The first reason is the fact that the thickness of the sapphire plates used is slightly thinner than the optimized one (Table \ref{tab:design}).
This causes an overall frequency shift to higher frequency of the modulation efficiency.
The second reason is caused by the large dips and oscillatory features caused by the air gaps, which also decreases the overall modulation efficiency when averaged over the band.
Particularly, this influence is large on the higher frequency bands.
The third reason is caused by the large phase variation on the higher and lower frequency bands (Table \ref{tab:phasediff}). 
Table \ref{tab:modeff_phase_pre} shows the prediction of the averaged modulation efficiencies and the maximum phase variation with reflection effects, no air gaps and the optimized thickness.
The band averaged modulation efficiencies at the highest and lower frequency band are less than 0.98.
This exception is due to the large phase variation in the frequency band.
The phase variation decreases in the amplitude of the modulated signal that is integrated in the band and the band averaged modulation efficiency.
Therefore, the averaged value becomes smaller, indicating that the phase variation is important.
The less phase variation within an observational band is also important to minimize the systematic effect, which results the susceptibility of the polarization angle sensitive orientation of a HWP polarimeter to the incident radiation spectrum and the system spectral shape.
Therefore, we study the further design optimization that minimizes the phase variation within the band while the modulation efficiency is kept high enough in 
preparation. 

\begin{table}[h]
\caption{The prediction of the averaged modulation efficiency and the maximum phase variation of the nine-layer AHWP with reflection effects, no air gaps, and optimized thickness. } 
\label{tab:modeff_phase_pre}
\begin{center}       
\begin{tabular}{|c|c|c|c|} 
	\hline
  	    band [GHz] & bandwidth [\%] & band averaged modulation efficiency & $\Delta \phi_{4}$ \\   \hline
	     40 & 30 & 0.969  &  $6.26^{\circ}$     \\  \hline
         50 & 30 & 0.977  &  $7.49^{\circ}$    \\  \hline
         60 & 23 & 0.982  &  $3.87^{\circ}$     \\  \hline
         68 & 23 & 0.991  &  $2.07^{\circ}$    \\  \hline
         78 & 23 & 0.995 &  $0.55^{\circ}$     \\  \hline
         89 & 23 & 0.995  &  $0.85^{\circ}$     \\  \hline
         100 & 23 & 0.989  &  $2.41^{\circ}$     \\  \hline
         119 & 30 & 0.990  &  $1.87^{\circ}$     \\  \hline
         140 & 30 & 0.993  &  $1.98^{\circ}$    \\ \hline
         166 & 30 & 0.993  &  $2.01^{\circ}$     \\ \hline
         195 & 30 & 0.990  &  $3.08^{\circ}$    \\ \hline
         235 & 30 & 0.938  &  $16.46^{\circ}$    \\ \hline
\end{tabular}
\end{center}
\end{table} 

\section{CONCLUSIONS}
\label{sec:conclusion}  
We design and evaluate the prototype of the nine-layer AHWP for use in CMB experiments.
We find the measurements in the modulation efficiency and the phase at the room temperature are in good agreement with the predictions.
Thus, we demonstrate our nine-layer AHWP to be broadband. 
However, we find small discrepancies between the measurements and the predictions. 
The primary contribution around the dips is attributed to the existence of air gaps between the stacked plates. 
The agreement gets even better when the effect is considered. 
We measure the incident angle dependence of the modulation efficiency and the phase in the incident angle range comparable to the field of view of LiteBIRD and the CMB telescope that observes small angle scales, about 10 degrees.
The dependence of the incident angle is found to be explained by the internal reflections in individual plates.
\par In many CMB experiments, the intensity of the observation signal is integrated by the detector in a specific bandwidth.
Therefore, we average the modulation efficiency over the bandwidth to obtain the value usable for the experiments.
From the evaluation using the band-averaged modulation efficiency, we find that the smaller the phase variation in the bandwidth, the larger the averaged modulation efficiency.
Therefore, the optimization to obtain higher modulation efficiency requires us to have uniform phase values in the bandwidth, which will be presented in future work.

\acknowledgments 
This paper based on a SPIE conference proceedings paper~\cite{kkomatsu_spie}.
This work is supported by JSPS KAKENHI Grant Number JP17H01125, JP15H05441, JP15H05891, JP17K14272, JP18J20148 and JSPS Core-to-Core Program, A. Advanced Research Networks, and the World Premier International Research Center Initiative (WPI Initiative), MEXT, Japan.
We would like to thank Dr. Samantha Stever for editorial suggestions to this paper.



\bibliography{report}   
\bibliographystyle{spiejour}   


\vspace{2ex}\noindent\textbf{First Author} is a PhD student at Okayama University. He received his BS and MS degrees in physics from Okayama University in 2016 and 2018, respectively. His current research interests the verification of inflation theory using B-mode polarization of CMB created by the primordial gravitational waves. Related to it, he is developing the polarization modulator of LiteBIRD.

\vspace{1ex}
\noindent Biographies and photographs of the other authors are not available.

\listoffigures
\listoftables

\end{document}